\documentclass[11pt,fleqn,twosides]{article}

\usepackage{mathrsfs}
\usepackage{latexsym}
\usepackage{amsmath,amsthm,amssymb,amsopn,amsfonts}
\usepackage{pstricks}
\usepackage{graphicx}

\usepackage{fancyhdr}
\usepackage{latexsym}
\fancypagestyle{plain}{ \fancyhead{} \fancyfoot{}

}

\newtheorem{theorem}{Theorem}
\newtheorem{lemma}[theorem]{Lemma}
\newtheorem{proposition}[theorem]{Proposition}

\newtheorem{corollary}[theorem]{Corollary}

\theoremstyle{remark}

\let\n\noindent

\begin{document}
\title{Hermite and Laguerre $\beta$-ensembles: asymptotic corrections to the eigenvalue density}

\author{Patrick  Desrosiers
\thanks{P.Desrosiers@ms.unimelb.edu.au} \cr
\emph{Department of Mathematics and Statistics},\cr University of
Melbourne,\cr Parkville, Victoria 3010,
 Australia.
\and Peter J.~Forrester
\thanks{P.Forrester@ms.unimelb.edu.au}\cr
\emph{Department of Mathematics and Statistics},\cr University of
Melbourne, \cr Parkville, Victoria 3010,
 Australia.}

\date{September 2005}

\maketitle

\begin{abstract}
We consider Hermite and Laguerre $\beta$-ensembles of large
$N\times N$ random matrices. For all $\beta$ even,
corrections to the limiting global density are obtained, and the limiting density at the soft edge is evaluated.
 We use the saddle point method on   multidimensional integral representations  of the density which are based on special realizations of the generalized
 (multivariate) classical orthogonal polynomials.
   The   corrections to the bulk density are oscillatory terms that depends on $\beta$.
   At the edges, the density can be expressed as a multiple integral of the Konstevich type which constitutes a $\beta$-deformation of  the Airy function.
     This allows us to obtain the main contribution to the soft edge density when the spectral parameter tends to
     $\pm\infty$.\\

{\footnotesize \noindent     {\bf 2000 MSC}: 15A52, 41A60, 33D52.\\
\noindent {\bf 2003 PACS}: 02.0Cw, 02.30Mv, 02.30Gp.}
\end{abstract}

\newpage

\tableofcontents

\section{Introduction
}
We deal with two families of $N\times N$ random
matrices: the Hermite and Laguerre  $\beta$-ensembles (for a review see
\cite{ForresterBook}). These ensembles possess an  eigenvalue
joint probability density function (p.d.f.)  of the form
\begin{equation}\label{pdf}
P_{N,\beta}(\mathbf{x})=\frac{1}{Z_{N}} e^{-\beta W
(\mathbf{x})},\qquad \mathbf{x}=(x_1,\ldots,x_N)\in I^N
,\end{equation} where $\beta$ is real and positive.   The support
$I$ of the eigenvalues in the Hermite and Laguerre cases are
respectively  $(-\infty,\infty)$ and $(0,\infty)$.  The
ensembles' names come from the fact that their p.d.f. generalize
the weight functions  related to the Hermite and Laguerre
polynomials; that is,
\begin{equation}\label{potentials}
  W (\mathbf{x})= \begin{cases}
    \displaystyle\frac{1}{2}\sum_{i=1}^Nx_i^2-\sum_{1\leq i<j\leq N}\ln |x_i-x_j|, & \text{Hermite}, \\
    \displaystyle \frac{1}{2}\sum_{i=1}^N x_i-\frac{a}{2}\sum_{i=1}^N\ln |x_i|-\sum_{1\leq i<j\leq N}\ln |x_i-x_j|,  &
    \text{Laguerre},
  \end{cases}
\end{equation} where $a$ is a real and nonnegative  parameter.
The normalization constants can be computed with the help of the
Selberg integrals:
\begin{equation}\label{ZN}{Z_{N}} = \begin{cases}
    \displaystyle G_{\beta,N}:=g_{\beta,N}\prod_{j=2}^N\frac{\Gamma(1+j\beta/2)}{\Gamma(1+\beta/2)}, & \text{Hermite}, \\
    \displaystyle W_{a,\beta,N}:=w_{a,\beta,N}\prod_{j=1}^N\frac{\Gamma(1+j\beta/2)\Gamma(1+(a+j-1)\beta/2)}{\Gamma(1+\beta/2)} & \text{Laguerre},   \end{cases}
\end{equation}
where $g_{\beta,N}=(2\pi)^{N/2}\beta^{-N(1/2+\beta (N-1)/4)}$ and $w_{a,\beta,N}=(2/\beta)^{N(a\beta/2+1+\beta (N-1)/2)}$.

For special values of the Dyson index $\beta$, we recover
classical random matrix ensembles (see e.g. \cite{ForresterBook,
Metha}). Indeed, the $\beta=1,2,$ and $4$
 Hermite ensembles are respectively  equivalent to the Gaussian orthogonal,
 unitary, and symplectic ensembles. The Laguerre
ensembles are similarly related to the real, complex and
quaternionic Wishart matrices. Recently, Dumitriu and Edelman
\cite{DumitriuMatrix} have constructed explicit random matrices
associated to the Hermite and Laguerre p.d.f. given in
Eq.\eqref{pdf}.  A generic random $N\times N$ matrix belonging to
the Hermite $\beta$-ensemble can be written as a tridiagonal
symmetric matrix:
\[H_\beta=\frac{1}{\sqrt{\beta}}\begin{pmatrix}
  \mathrm{N}[0,1] & {\chi }_{(N-1)\beta}&  & &   \\
{\chi }_{(N-1)\beta} & \mathrm{N}[0,1]&{\chi }_{(N-2)\beta}& &  &    \\
  & {\chi }_{(N-2)\beta} & \mathrm{N}[0,1] & {\chi }_{(N-3)\beta} &   \\
 &\;\ddots &\;\ddots & \;\ddots  & &  \\
    & &{\chi }_{2\beta} &  \mathrm{N}[0,1] & {\chi }_{\beta} \\
 &  && {\chi }_{\beta} &\mathrm{N}[0,1]
\end{pmatrix}.\]
This means that the $N$ diagonal elements and the $N-1$ subdiagonal
elements are mutually independent; the diagonal elements  are
normally  distributed (with mean zero and variance 1) while the
off-diagonal have a chi  distribution. Recall that the densities
associated to $\mathrm{N}[\mu,\sigma]$ and ${\chi }_{k}$ are
respectively $(2\pi\sigma^2)^{-1/2}e^{-(x-\mu)^2/(2\sigma^2)}$ and
$2x^{k-1}e^{-x^2}/\Gamma(k/2)$, where in the latter case $x>0$. Any
$N\times N$ matrix $L_\beta$ of the Laguerre $\beta$-ensemble also
has a tridiagonal form:
 $L_\beta=B_\beta^T B_\beta$, for some $N\times N$
matrix
\[ B_\beta=\frac{1}{\sqrt{\beta}}\begin{pmatrix}
{\chi }_{P\beta}&  {\chi }_{(N-1)\beta}& &   \\
&{\chi }_{(P-1)\beta} &{\chi }_{(N-2)\beta}&      \\
 &\, \ddots & \,\ddots & &  \\
&&{\chi }_{(P-N+1)\beta} &  {\chi }_{\beta} \\
\end{pmatrix},\qquad a=P-N+1-\frac{2}{\beta}.\]

In this article, we compute the density for large but finite
random matrices of the Hermite and Laguerre $\beta$-ensembles.
The density, or the marginal eigenvalue probability density, is
defined as follows:
\begin{equation}\label{defrho}
  \rho_{N,\beta}(x):=\frac{N}{Z_N}\int_{I^N}P_{N,\beta}(x_1,\ldots,x_N)\, dx_1\cdots
  dx_N.
\end{equation}
The quantity   $N^{-1}\rho_{N,\beta}(x)dx$ represents the probability
to have an eigenvalue in the interval $[x,x+dx]$.  The density
has two simple physical interpretations.

 First, we remark that the Hermite
p.d.f.\ is equivalent to the Boltzmann factor of a log-potential
Coulomb gas with particles of charge unity confined to the interval
$(-\sqrt{2N},\sqrt{2N})$ with neutralizing background charge density
$-(\sqrt{2N}/\pi)\sqrt{1-x^2/2N}$.  From this point of view, $Z_{N}$
(divided by $N!$) is simply the  canonical partition function at
inverse temperature $\beta$ and $ \rho_{N,\beta}(x)dx$ gives the
number  of charges present in the interval $[x,x+dx]$. This analogy
allows one to predict the  global density :
\begin{equation}\label{semicircle}\lim_{N\rightarrow\infty}\sqrt{\frac{2}{N}}\rho_{N,\beta}(\sqrt{2N}x)= \rho_{\mathrm{{W}}} (x):= \begin{cases}
    \displaystyle\frac{2}{\pi}\sqrt{1-x^2},&  -1< x< 1, \\
    \displaystyle 0,&
    |x|\geq 1.
  \end{cases}
\end{equation} This result is known as the Wigner  semicircle law.
For a finite matrix, we  expect that the scaled density is of order
one in the interval $(-\sqrt{2N},\sqrt{2N})$, the `bulk region' of
the mechanical problem,  while it decreases rapidly around
$\pm\sqrt{2N}$, called the `soft edges'.   A similar log-gas
construction is possible for the Laguerre case. One  expects  the
$c=1$ Mar\v{c}enko-Pastur law \cite{MarcenkoPastur}:
\begin{equation}\label{semicircleL}
\lim_{N\rightarrow\infty}4 \rho_{N,\beta}(4N
x)=\rho_\mathrm{MP}(x):=\begin{cases}
    \displaystyle\frac{2}{\pi}\sqrt{\frac{1}{x}-1},&  0< x< 1, \\
    \displaystyle 0,&
    x\geq 1.
  \end{cases}
\end{equation}We see that, in the Laguerre case, the `bulk' is $(0,4N)$ while the
`soft edge'   is the point $4N$.  The origin is referred as the
`hard edge' of the support because the eigenvalues are constrained
to be positive. The  predictions   given in Eqs \eqref{semicircle}
and \eqref{semicircleL}  have been confirmed in
\cite{Baker,ForresterLaguerre}.  The asymptotic analysis used in
these references constitutes the starting point for the study of the
higher expansions to be undertaken in the present work.

Second, their is a deep connection between the $\beta$-ensembles
and some integrable quantum mechanical $N$-body problems on the
line, known as the Calogero-Moser-Sutherland (CMS) models
(a good reference is \cite{Perelomov}).  The Hermite p.d.f. is in fact
 the ground state wave functions squared of the (rational) $A_{N-1}$  CMS model, whose  Hamiltonian is \[H^{\mathrm{(H)}}=-\sum_{i=1}^N
\frac{\partial^2}{\partial x_i^2}+\frac{\beta^2}{4}\sum_{i=1}^Nx_i^2
+\frac{\beta(\beta-2)}{2}\sum_{1\leq i<j\leq N}\frac{1}{(x_i-x_j)^2}
,\]for $x_j\in (-\infty,\infty)$. The Laguerre p.d.f. is the ground
state squared of the Hamiltonian of the $B_{N}$ CMS model, which can
be expressed as follows:
\[\begin{split}H^{\mathrm{(L)}}=&-2\sum_{i=1}^N
\left( 2x_i\frac{\partial^2}{\partial
x_i^2}+\frac{\partial}{\partial x_i}\right)\\&
+\frac{1}{4}\sum_{i=1}^N\left(a\beta(a\beta-2)\frac{1}{x_i}+\beta^2x_i\right)
+{\beta(\beta-2)}\sum_{1\leq i<j\leq
N}\frac{x_i+x_j}{(x_i-x_j)^2},\end{split}\] where $x_j\in
(0,\infty)$.  It has been shown in \cite{Baker} (see also
\cite{vandiejen}) that the eigenfunctions of the conjugated
Schr\"odinger operators $e^{\beta W/2}H^{\mathrm{(H)}}e^{-\beta
W/2}$ and $e^{\beta W/2}H^{\mathrm{(L)}}e^{-\beta W/2}$ are
respectively the generalized (or multivariate) Hermite and Laguerre
polynomials, previously introduced by Lassalle in \cite{LassalleH,
LassalleL}.   In the context of CMS models, the global density can
be seen as  the ground state expectation value of the density
operator $\hat\rho(x)=\sum_{j=1,\ldots,N}\delta(x-x_j)$, also known
as the one-point  function.

The relation between the CMS models and the generalized classical orthogonal polynomials furnishes, when $\beta$ is an even integer,
 new integral representations of the global density that suits perfectly for asymptotic  analysis.  Let us be more explicit.
The  definition of the density given in \eqref{defrho} contains $N$ integrals; considering $N$ large does not simplify the calculation.
 On the other hand,  it has been noticed in \cite{Baker,ForresterLaguerre} that  the density is a particular Hermite (or Laguerre)
  polynomial, characterized  by a
partition $\lambda=((N-1)^\beta)$ and evaluated at
$x_1=\ldots=x_\beta=x$ (see below). Using the work of Kaneko
\cite{Kaneko} and Yan \cite{Yan}, one then can shows that the
density is proportional to the following $\beta$-dimensional
integral:
\begin{equation}\label{Rdef}R_{N,\beta}(x):=\int_\mathcal{C }du_1\, e^{Nf(u_1,x)}\cdots\int_\mathcal{C}
du_\beta\, e^{Nf(u_\beta,x)}\, \prod_{1\leq j<k\leq
\beta}|u_j-u_k|^{4/\beta}, \end{equation} for a particular  contour
$\mathcal{C}$ and  function $f(u,x)$.

In the following sections, we apply the steepest descent method
\cite{Olver,Wong} to integrals of the type  \eqref{Rdef}.   We
obtain expressions for the density in the bulk and at the soft
edge that are valid for every $\beta\in 2 \mathbb{N}$. Of course,
these results  generalize many known result obtained for
$\beta=2$ and $4$.  We mention in particular two recent
publications in which asymptotic corrections to the global
density have been obtained: 1)
 Kalisch and Braak \cite{Kalisch} for some ensembles, including
the Gaussian unitary and symplectic ensembles (work based on the
supersymmetric method); 2) Garoni, Frankel and Forrester
\cite{Garoni} for the Laguerre and Gaussian unitary ensembles
(calculations using the theory of orthogonal polynomials). Also, the
preprint \cite{Garoni2} of Forrester, Frankel and Garoni addresses
the Laguerre and Gaussian ensembles with orthogonal and symplectic
symmetry. All studies show that these approximate expressions of the
global density are very accurate, even for $N=10$, say (for
instance, see Fig.~1 and Fig.~2 in \cite{Garoni}).  We finally point
out that an asymptotic formula for the density in the Hermite
$\beta$-ensemble has been considered in a different context:
Johansson \cite{Johansson} has studied a smoothed (macroscopic)
density and has derived corrections of order $1/N$ to
Eq.~\eqref{semicircle}. However, contrary to the asymptotic formula
obtained here, the large $N$ expansion given in \cite{Johansson}
does not contain oscillatory (microscopic) terms.

The  article is organized as follows.  In Section 2, we
review the exact expressions of the densities in terms of the
generalized Hermite and Laguerre polynomials.  In Section 3, we
derive the first oscillatory corrections to the global densities
\eqref{semicircle} and \eqref{semicircleL}.   These
approximations are also compared to the exact densities given in
Section 2.  The asymptotic densities evaluated about the soft
edges of the spectrum are obtained in Section 4; they are
expressed in terms of Kontsevich type integrals.  The behavior of
the latter when the spectral parameter is large is studied in
Section 4.  In the last section, we finally summarize the
principal results and discuss the generalization of some of our results to general $\beta$.

\section{Exact expressions of the density}

As previously mentioned, the density in the Hermite and Laguerre ensembles can
be written as particular generalized Hermite and Laguerre
polynomials \cite{Baker}.  These polynomials are   symmetric, so we
can write them as a linear combination of monomial symmetric
functions
\[m_\lambda(x_1,\ldots,x_N):=x_1^{\lambda_1}\cdots
x_N^{\lambda_N}+\mbox{distinct permutations},\]where
$\lambda=(\lambda_1,\ldots,\lambda_N)$ is a partition of weight
$|\lambda|=\sum_{i=1}^N\lambda_i$.  It is convenient to introduce
another basis of the algebra of symmetric polynomials, namely,
the (monic) Jack polynomials $\bar{J}^{(\alpha)}_\lambda$.  They
constitute the only basis such that\begin{gather*}
\bar{J}_\lambda^{(\alpha)}(x_1,\ldots,x_N)=m_\lambda(x_1,\ldots,x_N)+\sum_{\mu<\lambda}a_{\lambda\mu}(\alpha)\,m_\mu(x_1,\ldots,x_N)\qquad \mbox{(triangularity)}\\
D_2^{(\alpha)}\,
 \bar{J}_\lambda^{(\alpha)}(x_1,\ldots,x_N)=\epsilon_\lambda(\alpha)
 \,\bar{J}_\lambda^{(\alpha)}(x_1,\ldots,x_N)\qquad \mbox{(eigenfunction)}\end{gather*}
 for some eigenvalue
 $\epsilon_\lambda(\alpha)$.  In the last equations, $\mu<\lambda$
 means that $\sum_{j=1}^k\mu_i\leq\sum_{j=1}^k\lambda_i$ for all $k$ when $|\mu|=|\lambda|$ but $\mu\neq\lambda$, while $D_2^{(\alpha)}$ is a
 particular differential operator that can be defined via
 \[D_k^{(\alpha)}:=\sum_{i=1}^Nx_i^k\frac{\partial^2}{\partial x_i^2}
 +\frac{2}{\alpha}\sum_{1\leq i<j\leq N}\frac{1}{x_i-x_j}\left(x_i^k\frac{\partial}{\partial x_i}-x_j^k\frac{\partial}{\partial
 x_j}\right).\]

The generalized
 Hermite polynomials, denoted  by $\bar{H}_\lambda(x_1,\ldots,x_N;\alpha)$,  are the
 only symmetric polynomials obeying to
\begin{gather*}
\bar{H}_\lambda
(x_1,\ldots,x_N;\alpha)=\bar{J}^{(\alpha)}_\lambda(x_1,\ldots,x_N)+
 \sum_{\substack{|\mu|=|\lambda|-2n\\n=1,2,\ldots,\lfloor|\lambda/2|\rfloor}}b_{\lambda\mu}(\alpha,N)\,\bar{J}^{(\alpha)}_\mu(x_1,\ldots,x_N),\\
(D_0^{(\alpha)}-2E_1)\,
 \bar{H}_\lambda(x_1,\ldots,x_N;\alpha)=-2|\lambda|\,
 \bar{H}_\lambda(x_1,\ldots,x_N;\alpha),\end{gather*}
 where
\[E_k:=\sum_{i=1}^Nx_i^k\frac{\partial}{\partial
 x_i}.\] Let us point out that
 \[D_0^{(\alpha)}-2E_1=-\frac{2}{\beta}e^{\beta
W/2}\,H^{\mathrm{(H)}}\,e^{-\beta W/2}+\mathrm{cst},\qquad
\beta=\frac{2}{\alpha} ,\]
 where $H^{\mathrm{(H)}}$ is
  the CMS Hamiltonian defined in Section 1.  On can show that
  \begin{equation}\label{explicitHermite}
  \bar{H}_\lambda(x_1,\ldots,x_N;\alpha)=\exp\left(-\frac{1}{4}D_0^{(\alpha)}\right)\bar{J}^{(\alpha)}_\lambda
  \end{equation}

  Similarly to the Hermite case, the generalized
 Laguerre polynomials, written $\bar{L}^{\nu}_\lambda(x_1,\ldots,x_N;\alpha) $,  are the
unique  symmetric polynomials satisfying
\begin{gather*}
\bar{L}^\nu_\lambda
(x_1,\ldots,x_N;\alpha)=\bar{J}^{(\alpha)}_\lambda(x_1,\ldots,x_N)+
 \sum_{\substack{|\mu|=|\lambda|-n\\ n=1,2,\ldots, |\lambda| }}c_{\lambda\mu}(\alpha,\nu,N)\,\bar{J}^{(\alpha)}_\mu(x_1,\ldots,x_N),\\
(D_1^{(\alpha)}-E_1+(\nu+1)E_0 )\,
 \bar{L}^\nu_\lambda(x_1,\ldots,x_N;\alpha)=-|\lambda|\,
 \bar{L}^\nu_\lambda(x_1,\ldots,x_N;\alpha).\end{gather*}
 The latter eigenvalue problem is related to a CMS model:
  \[D_1^{(\alpha)}-E_1+(\nu+1)E_0=-\frac{1}{2\beta} e^{\beta
W/2}\,H^{\mathrm{(L)}}\,e^{-\beta W/2}+\mathrm{cst},\qquad \beta=\frac{2}{\alpha} ,\qquad \nu=\frac{\beta a-1}{2}.  \]  The following
formula furnishes a way to  compute the generalized Laguerre
polynomials:
\begin{equation}\label{explicitLaguerre}
  \bar{L}^{\nu}_\lambda(x_1,\ldots,x_N;\alpha)=\exp\left(-D_1^{(\alpha)}-(\nu+1)E_0\right)\bar{J}^{(\alpha)}_\lambda(x_1,\ldots,x_N).
\end{equation}

When $\beta$ is an even integer, the density in the Hermite and
Laguerre  ensembles can be respectively written as a particular
Hermite and Laguerre polynomial; explicitly,
\begin{equation}\label{rhoexact}{\small\rho_{N,\beta}(x)=\begin{cases}\displaystyle
N\frac{G_{\beta,N-1}}{G_{\beta,N}}e^{-\beta x^2/2}
\bar{H}_{((N-1)^\beta)}(x_1,\ldots,x_\beta;\beta/2)\big|_{x_1=\ldots=x_\beta=x},&\mbox{Hermite},\\
\displaystyle N\frac{W_{a,\beta,N-1}}{W_{a,\beta,N}}x^{a\beta/2}
e^{-\beta x/2}
\bar{L}^{a-1+2/\beta}_{((N-1)^\beta)}(x_1,\ldots,x_\beta;\beta/2)\big|_{x_1=\ldots=x_\beta=x},&\mbox{Laguerre},
\end{cases}}\end{equation} where we have used the convention $(n^k)=\overbrace{(n,\ldots,n)}^k$.
Eqs \eqref{explicitHermite} and \eqref{explicitLaguerre},
together with  the fact that
$\bar{J}^{(\alpha)}_\lambda(x_1,\ldots,x_k)=x_1^n\cdots x_k^n$
when $\lambda=(n^k)$,  readily imply the following exact expressions of the density:
\begin{equation}\label{ExactDensityH}\rho_{N,\beta}(x)=N\frac{G_{\beta,N-1}}{G_{\beta,N}}e^{-\beta x^2/2}
 \sum_{n=0}^{\beta(N-1)/2}\left[\frac{(-1)^n}{4^n
  n!}\left(D_0^{(\beta/2)}\right)^n x_1^{N-1}\cdots
  x_\beta^{N-1}\right]_{x_1=\ldots=x_\beta=x}\end{equation} in the Hermite
  case, and
\begin{multline}\label{ExactDensityL}\rho_{N,\beta}(x)=N\frac{W_{a,\beta,N-1}}{W_{a,\beta,N}}x^{a\beta/2}e^{-\beta x/2}\\
\times\sum_{n=0}^{\beta(N-1)}\left[\frac{(-1)^n}{
  n!}\left(D_1^{(\beta/2)}+(a-2/\beta)E_0\right)^n x_1^{N-1}\cdots
  x_\beta^{N-1}\right]_{x_1=\ldots=x_\beta=x},\end{multline} in the Laguerre
  case.

We will use the two latter formulas to compare the exact and
asymptotic expressions of the density.  Note however that Eqs
\eqref{ExactDensityH} and \eqref{ExactDensityL} are computable only
when both $N$ and $\beta$  are small (i.e., for small partitions in
Eq.~\eqref{rhoexact}). There exist other methods that allow to
calculate the multivariate classical polynomials (see for instance
\cite{DumitriuMOPS}), but they  suffer from the same restrictions.  As
shown in the next section, the asymptotic expressions provide a more
tractable way to determine the density when $N$ is large.

\section{Density in the bulk}

  In this section, we obtain oscillatory corrections to the global densities  \eqref{semicircle} and \eqref{semicircleL}.
  This is achieved by deforming the contours of integration $\mathcal{C}$ in \eqref{Rdef} in such a way that they pass through
  the saddle points of the function $f(u,x)$.  In both the Hermite and the Laguerre cases, the function $f(u,x)$ has two simple saddle points
  in the complex $u$-plane, called $u_+$ and $u_-$.
  All oscillatory terms can be seen as  combinatorial corrections: the global density is  recovered when $\beta/2$ variables go through  $u_+$ while
  the remaining $\beta/2$ variables go through $u_-$; the dominant  oscillatory term comes from the integration of $\beta/2+1$ variables through  $u_+$ and $\beta/2-1$ variables through $u_-$ and conversely; the second oscillatory term comes the integration of $\beta/2+2$ variables through $u_+$ and $\beta/2-2$ variables through $u_-$ and conversely; and so on.

\subsection{Hermite case}

Before considering explicitly the density in the Hermite ensemble,
we prove two technical results associated to the asymptotics of the
integral \eqref{Rdef}.  In the following lines, we suppose that
$e^{Nf(u,x)}$ is analytic everywhere in the finite complex
$u$-plane, except possibly at a pole depending on $x$, and that
$\mathcal{C}$ is the real interval $(-\infty, \infty)$.

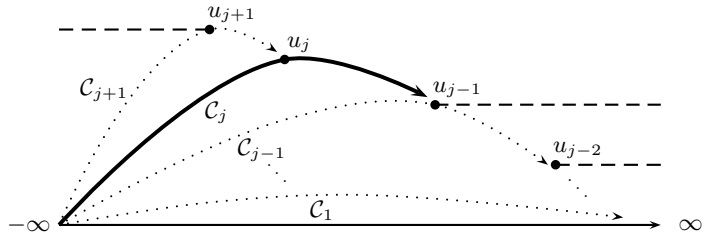
\begin{figure}[h]\caption{{\footnotesize New contours $\{\mathcal{C}_1,\ldots,\mathcal{C}_n\}$ in the complex $u_j$-plane.}}\label{newpaths}
\begin{center}
 \begin{pspicture}(0,0)(10,3)
\rput(4.5,0.2){{\footnotesize $\mathcal{C}_1$}}
\rput(3.9,0.8){{\tiny$\ddots$}}
      \rput(3.7,1){{\footnotesize $\mathcal{C}_{j-1}$}}
       \rput(3.1,1.5){{\footnotesize $\mathcal{C}_{j}$}}
        \rput(1.6,1.8){{\footnotesize $\mathcal{C}_{j+1}$}}
        \rput(7.9,0.6){{\tiny$\ddots$}}
 \psline{->}(1,0)(9,0)
  \rput(0.6,0){{\footnotesize $-\infty$}}
  \rput(9.4,0){{\footnotesize $\infty$}}
  \psline[linestyle=dashed]{-}(1,2.6)(3,2.6)
  \psline[linestyle=dashed]{-}(6,1.6)(9,1.6)
   \psline[linestyle=dashed]{-}(7.6,0.8)(9,0.8)
   \pscurve[linestyle=dotted]{->}(1.0,0.0)(3,2.6)(3.9,2.3)
  \pscurve[linestyle=dotted]{->}(1.0,0.0)(4,1.4)(6,1.6)(7.5,0.8)
  \pscurve[linestyle=dotted]{->}(1.0,0.0)(3,0.3)(5,0.4)(8.5,0.1)
  \pscurve[linewidth=1.5pt]{->}(1.0,0.0)(4,2.2)(5.9,1.7)
  \psdots*[dotstyle=*](6,1.6)(3,2.6)(4,2.2)(7.6,0.8)
  \rput(3.3,2.8){{\footnotesize $u_{j+1}$}}
  \rput(4.2,2.4){{\footnotesize $u_{j}$}}
   \rput(6.3,1.8){{\footnotesize $u_{j-1}$}}
     \rput(7.9,1){{\footnotesize $u_{j-2}$}}
 \end{pspicture}
\end{center}
\end{figure}

The method of steepest descent requires that the integrand of
\eqref{Rdef} should be analytic. This means in particular that the
absolute values must be removed. Such an  operation is realized in
the following lemma; it is possible when the line integration
$\mathcal{C}$ of the variable $u_j$ is deformed into an appropriate
complex path $\mathcal{C}_j$. Acceptable contours are given in
Fig.~\ref{newpaths}.  Other appropriate contours are obtained by
making a reflection of the picture with respect to the real axis.
Note that the dashed lines stand for (movable) branch cuts.  We
stress that $u_j$'s contour starts at $-\infty$ and ends at the
complex variable $u_{j-1}$. Only the path of $u_1$ (the last
variable to be integrated) ends on the real axis.

\begin{lemma}\label{orderedingH}Let $\{\mathcal{C}_j\}$ be a set of non-intersecting contours such
that $\mathcal{C}_1$ is a simple contour going from $-\infty$ to
$\infty$ and such that $\mathcal{C}_j$ goes from $-\infty$ to
$u_{j-1}$ for all  $j=2,\ldots, n$ (see Fig.~\ref{newpaths}). Then
\begin{multline*}
\int_{-\infty}^\infty du_1\cdots\int_{-\infty}^\infty du_n\prod_{i=1}^n  e^{Nf(u_i,x)} \prod_{1\leq j<k\leq
n}|u_j-u_k|^{4/\beta}= \\
\ n!\int_{\mathcal{C}_1}du_1\cdots \int_{\mathcal{C}_n}du_n
\prod_{i=1}^n e^{Nf(u_i,x)} \prod_{1\leq j<k\leq
n}(u_j-u_k)^{4/\beta},\end{multline*} where $-\pi<\mathrm{arg}\,
u_j\leq \pi$ and where $\mathrm{arg}\,(u_i-u_j)^{4/\beta}=0$ when
$u_i,u_j\in\mathbb{R}$ but $u_i>u_j$.\footnote{ When $n\neq \beta$ and $\beta\not\in\mathbb{N}$ in Lemma \ref{orderedingH},  other  integral representations in which all  variables go from $-\infty$ to $\infty$ are possible \cite{Dotsenko}.}
\end{lemma}
\begin{proof}Using the invariance of the lefthand side of the previous equation under any permutation, we
immediately see that it is equivalent to the following ordered
integrals:
\[ n!\int_{-\infty}^\infty du_1\,
e^{Nf(u_1,x)}\int_{-\infty}^{u_{1}} du_2\,
e^{Nf(u_2,x)}\cdots\int_{-\infty}^{u_{n-1}} du_n\, e^{Nf(u_n,x)}\,
\prod_{1\leq j<k\leq n}(u_j-u_k)^{4/\beta}\, .\]These integrals only
contain analytic functions, so  we can use Cauchy's Theorem. This
implies that the contour of $u_1$ can be deformed into  any simple
curve starting at  $-\infty$ and stopping at $\infty$. For the
remaining variables, any nonintersecting contour of integration in
the complex plane which starts at $-\infty$,  doesn't cross the
branch cuts coming coming from the multivaluedness of the integrand,
and complies with the ordering of the variables can be chosen.
\end{proof}

We are now in position to analyse  \eqref{Rdef} when $N$ is large.
Recall that the basic idea of the steepest descent method is to
choose a path for which the decrease of $f(u,x)$ in maximum. In
particular, this means that the contour must pass through the
saddle points.  In the bulk case, $f$ has two simple saddle points
$u_\pm$; that is,
\begin{equation}\label{2sp}\frac{\partial}{\partial
u}f(u,x)\big|_{u_\pm}=0,\qquad \frac{\partial^2}{\partial
u^2}f(u,x)\big|_{u_\pm}=R e^{\mathrm{i}\phi_\pm}, \quad
R>0.\end{equation}The directions of steepest descent at these
points, denoted $\theta_\pm$, are such that
$\cos(2\theta_\pm+\phi_\pm)=-1$ and
$\sin(2\theta_\pm+\phi_\pm)=0$,
so
\begin{equation}\label{thetapm}\theta_\pm=\frac{\pi-\phi_\pm}{2}\;
(\mathrm{mod}\, \pi),\qquad -\pi< \theta_\pm\leq\pi.\end{equation}

\begin{proposition} \label{propRH} Let $f(u,x)$ be a function that satisfies
 Eqs \eqref{2sp} and \eqref{thetapm}.  Let also
$f_\pm=f(u_\pm,x)$.  Suppose moreover that the saddle points are
such that $ \Re(u_-)<\Re(u_+)$.  Then,
\begin{multline*}R_{N,\beta}(x)=\binom{ \beta}{ \beta/2}(\Gamma_{\beta/2,\beta})^2(u_+-u_-)^\beta\left(\frac{2}{NR}\right)^{\beta-1}\\
\times e^{\beta N(f_++f_-)/2}e^{\mathrm{i}(\beta-1)(\theta_++\theta_-)}
\left[r_{N,\beta}(x)+\mathrm{O}\left(\frac{1}{N}\right)\right],\end{multline*}
where
\begin{multline*}
 r_{N,\beta}(x)=1 +2 \sum_{k=1}^{\lfloor \sqrt{\beta/2}\rfloor}\left[\left(\prod_{j=1}^k\frac{\Gamma(1+2j/\beta)}{\Gamma(1+2(j-k)/\beta)}\right)\right.\\
\times\left.\frac{e^{\mathrm{i}2k^2(\theta_++\theta_-)/\beta}}{(u_+-u_-)^{4k^2/\beta}(NR)^{2k^2/\beta}}
\cos\Big(-\mathrm{i}k N
(f_+-f_-)+k(\theta_+-\theta_-)(3-2/\beta)\Big)\right]\end{multline*}
and
\begin{equation}\label{defGamma}
\Gamma_{n,\beta}:=\int_{-\infty}^\infty du_1\cdots
\int_{-\infty}^\infty du_n \prod_{i=1}^n  e^{-u_i^2}\,
\prod_{1\leq j<k\leq n}|u_j-u_k|^{4/\beta}
=\frac{\pi^{n/2}}{2^{
n(n-1)/\beta}}\prod_{j=2}^n
\frac{\Gamma(1+2j/\beta)}{\Gamma(1+2/\beta)}.\end{equation}
\end{proposition}
\begin{proof}We first apply Lemma \ref{orderedingH} to the expression \eqref{Rdef}.  Then, the contours
$\mathcal{C}_j$ are deformed into steepest descent contours
$\mathcal{S}_j=\mathcal{S}_j^-\cup\mathcal{S}_j^+$ passing through
the saddle points $u_\pm$. Close to these points, the contours are
parametrized as follows:
\[u_j=u_\pm+t_je^{\mathrm{i}\theta_\pm},\quad-\pi<\mathrm{arg}\,u_j\leq\pi,
\quad\mbox{on}\quad\mathcal{S}_j^\pm,\]  where the angles of steepest
descent are given \eqref{thetapm} and where $t_j\in (-\tau,\tau)$
{for some} $\tau>0$. Moreover, we impose $t_i>t_j$ for $i<j$ in
order to guarantee $\Re (u_{i})>\Re (u_{j})$ for $i<j$. Setting
\[y_j=\sqrt{\frac{NR}{2}}t_j \]  we obtain
\[y_i\in(-\infty,\infty)\quad\mbox{and}\quad
f(u_j,x)=Nf_\pm-y_j^2+\mathrm{O}(1/\sqrt{N}) \]as
$N\rightarrow\infty$.  When both $u_j$ and $u_k$ are close to the
same saddle point $u_\pm$,
\[(u_j-u_k)^{4/\beta}=\left(\frac{2}{NR}\right)^{2/\beta}
e^{\mathrm{i}4\theta_\pm/\beta}(y_j-y_k)^{4/\beta}\,.\] When $u_j$
is on $\mathcal{S}_j^+$ while $u_k$ is on $\mathcal{S}_j^-$, we
have
\[(u_j-u_k)^{4/\beta}=(u_+-u_-)^{4/\beta}+\mathrm{O}(1/\sqrt{N})\,.\]

We now return to the expression of $R_{N,\beta}(x)$ by
considering the steepest descent paths:
\begin{multline*}
R_{N,\beta}(x)=\beta!\sum_{n=0}^{\beta}
\left(\int_{\mathcal{S}_j^+}du_1\cdots\int_{\mathcal{S}_j^+}du_n\right.
\\ \times \left.\int_{\mathcal{S}_j^-}
du_{n+1}\cdots\int_{\mathcal{S}_j^-} du_{\beta}\prod_{j=1}^
ne^{Nf(u_j,x)}\prod_{1\leq j<k\leq
\beta}(u_j-u_k)^{4/\beta}\right).\end{multline*}
 In terms of the new variables
introduced above, the righthand side of the previous equation
becomes \begin{multline*}
\beta!\sum_{n=0}^\beta S_{n}\left[\int_{-\infty}^{\infty}
dy_1\int_{-\infty}^{y_1} dy_2\cdots\int_{-\infty}^{y_{n-1}}
dy_n\prod_{j=1}^n
e^{-y_j^2}\left(1+\mathrm{O}\left(\frac{1}{\sqrt{N}}\right)\right)\prod_{1\leq
p<q\leq
n}(y_p-y_q)^{4/\beta}\right.\\
\left.\times\int_{-\infty}^{\infty}
dy_{n+1}\int_{-\infty}^{y_{n+1}}
dy_{n+2}\cdots\int_{-\infty}^{y_{\beta-1}}
dy_\beta\prod_{k=n+1}^\beta
e^{-y_k^2}\left(1+\mathrm{O}\left(\frac{1}{\sqrt{N}}\right)\right)\prod_{n+1\leq
r<s\leq \beta}(y_r-y_s)^{4/\beta}\right]\,,\end{multline*}where
\begin{multline*}S_{n}=(u_+-u_-)^{4
n(\beta-n)/\beta}\left(\frac{2}{NR}\right)^{3\beta/2-1-2n(\beta-n)/\beta}
\\ \times e^{Nn f_+ +N(\beta-n)f_-} e^{\mathrm{i}\theta_+(n+
2n(n-1)/\beta)}e^{\mathrm{i}\theta_-(\beta-n+2
(\beta-n)(\beta-n-1)/\beta)} .\end{multline*}

The Gaussian terms are in fact ordered versions of the functions
$\Gamma_{n,\beta}$, introduced in Eq.\eqref{defGamma}. We again
use   Lemma \ref{orderedingH} and get
\begin{equation*}\label{bigterm}
\begin{split} R_{N,\beta}(x)&= \sum_{n=0}^\beta \binom{\beta}{n}
\left[ S_{n}
\Gamma_{n,\beta}\Gamma_{\beta-n,\beta}+\mathrm{O}\left(\frac{1}{N}\right)\right]\\
&= \binom{\beta}{\beta/2}\left(\Gamma_{ \beta/2,\beta}\right)^2S_{ \beta/2}\left[r_{N,\beta}(x)+\mathrm{O}\left(\frac{1}{N}\right)\right],\end{split}
\end{equation*}
where
 \begin{equation} \label{rNbetacomplete}r_{N,\beta}(x)=1+\sum_{k=1}^{\beta/2}
\binom{\beta}{ \beta/2+k} \binom{ \beta}{
\beta/2}^{-1}\frac{\Gamma_{ \beta/2+k,\beta}\Gamma_{
\beta/2-k,\beta}}{\Gamma_{ \beta/2,\beta}\Gamma_{
\beta/2,\beta}}\left[\frac{ S_{ \beta/2+k}+S_{
\beta/2-k}}{S_{\beta/2}}
+\mathrm{O}\left(\frac{1}{N}\right)\right].\end{equation}

Note that the order of the `analytic corrections' (i.e., coming from
the  expansion of $f(u_j,x)$ as a polynomial in  $y_j$ of degree
superior than 2) is now $1/N$ rather than $1/\sqrt{N}$. This can be
explained by using the theory of the generalized Hermite polynomials
\cite{Baker}: only symmetric polynomials $p(y_1,\ldots,y_n)$ of
degree even may have a nonzero contribution to
\[\int_{-\infty}^\infty dy_1\cdots\int_{-\infty}^\infty dy_n
\prod_{i=1}^ne^{-y_i^2}\prod_{1\leq j<k\leq
n}|y_j-y_k|^{4/\beta}p(y_1,\ldots,y_n).\]In our case, the
$\mathrm{O}(1/\sqrt{N})$ terms are symmetric polynomials in $y_j$ of
degree one and three, so they don't contribute in the expression of
$R_{N,\beta}(x)$.

Note also that $(S_{ \beta/2+k}+S_{ \beta/2-k})/{S_{\beta/2}}$ is of
order $1/N^{2k^2/\beta}$.  Hence, the `combinatorial terms' with
$k>\lfloor \sqrt{\beta/2}\rfloor$ are smaller than the `analytic
corrections' of order $1/N$; thus  we must truncate
Eq.~\eqref{rNbetacomplete} as follows:
 \begin{equation*} r_{N,\beta}(x)=1+\sum_{k=1}^{\lfloor \sqrt{\beta/2}\rfloor}
\binom{\beta}{ \beta/2+k} \binom{ \beta}{
\beta/2}^{-1}\frac{\Gamma_{ \beta/2+k,\beta}\Gamma_{
\beta/2-k,\beta}}{\Gamma_{ \beta/2,\beta}\Gamma_{
\beta/2,\beta}}\left[\frac{ S_{ \beta/2+k}+S_{
\beta/2-k}}{S_{\beta/2}}\right].\end{equation*}This expression
contains only `combinatorial corrections'.

Finally,  one readily shows that
\[\binom{\beta}{ \beta/2+k}
\binom{ \beta}{  \beta/2}^{-1}\frac{\Gamma_{ \beta/2+k,\beta}\Gamma_{  \beta/2-k,\beta}}{\Gamma_{ \beta/2,\beta}\Gamma_{  \beta/2,\beta}}
=\frac{1}{2^{2k^2/\beta}}\prod_{j=1}^k\frac{\Gamma(1+2j/\beta)}{\Gamma(1+2(j-k)/\beta)}\]and
the proof is complete.\end{proof}

The explicit link between the Hermite  density and the
integral of the type \eqref{Rdef} has been obtained in
\cite{Baker}.  It reads:
\begin{equation}\label{rhoRHermite}\rho_{N,\beta}(\sqrt{2N}x)=\frac{1}{2}
\frac{G_{\beta,N-1}}{G_{\beta,N} \Gamma_{\beta,\beta}}(2N)^{\beta
N/2+\beta}e^{-\beta N x^2}R_{N,\beta}(x) \end{equation}if
\begin{equation}\label{fhermite}f(u,x)=-2u^2+\ln
(\mathrm{i}u+x)-\frac{1}{N}\ln (\mathrm{i}u+x).\end{equation} Up to additive terms of order $1/N$, the
latter  function has two saddle points
\[u_\pm=\frac{1}{2}(\mathrm{i}x\pm\sqrt{1-x^2}).\] We see that $\Re(u_+)>\Re(u_-)$ only if $-1<x<1$.
  According to the notation used in Eqs \eqref{2sp} and \eqref{thetapm}, we have
\[f_\pm=-\frac{1}{2}-\frac{N-1}{N}\ln
2+x^2\pm\mathrm{i}\left(\frac{N-1}{N}\arccos
x-x\sqrt{1-x^2}\right)
\]and \[Re^{\mathrm{i}\phi_\pm}=8\sqrt{1-x^2} e^{\mathrm{i}(\pi-\arcsin x)}, \]
where we have made use of
\[\arcsin x=-\mathrm{i}\ln
(\mathrm{i}x+\sqrt{1-x^2})= {\pi}/{2}-\arccos x.\] Note that the
inverse trigonometric functions are defined on their principal
branch; that is,   $\arcsin x\,:\,
[-1,1]\longrightarrow[-\pi/2,\pi/2]$ and $\arccos x\,:\,
[-1,1]\longrightarrow[\pi,0]$.  Thus the angles of steepest descent
are
\[\theta_{\pm}=\mp\frac{1}{2}\arcsin x\, .\]
Stirling's approximation,
 \[\Gamma(y+z)=\sqrt{2\pi}e^{-z}z^{y+z-1/2}\left[1+\mathrm{O}\left(\frac{1}{z}\right)\right]\quad \mbox{when}\quad z\rightarrow\infty, \]
 immediately implies
  \[\frac{G_{\beta,N-1}}{G_{\beta,N}} = \frac{2^{\beta N/2-1/2}\beta^{\beta N/2-\beta}e^{\beta N/2}}{\pi N^{\beta N/2+1/2}}\left[1+\mathrm{O}\left(\frac{1}{N}\right)\right].\]
The substitution of the above results in Proposition \ref{propRH}
provide  the sought asymptotic  corrections to the  global density.

\begin{corollary}
Let $-1<x<1$ and let $P_\mathrm{W}(x)$ denote the (cumulative)
probability distribution associated to the semicircle law given in
Eq.~\eqref{semicircle}; i.e.,
\begin{equation}\label{PW} P_\mathrm{W}(x)=\int_{-1}^x \rho_\mathrm{W}(t) dt=1+\frac{x}{2}\rho_\mathrm{W}(x) -\frac{1}{\pi}\arccos x\,
.\end{equation}Then
\[\sqrt{\frac{2}{N}}\rho_{N,\beta}(\sqrt{2N}x)=\rho_\mathrm{W}(x) r_{N,\beta}(x)+\mathrm{O}\left(\frac{1}{N}\right), \]
where
\begin{multline}\label{ADensityH}
 r_{N,\beta}(x)= 1+2\sum_{k=1}^{\lfloor\sqrt{\beta/2}\rfloor}\frac{(-1)^k}{(\pi^3\rho_\mathrm{W}(x)^3 N)^{2k^2/\beta}}\\ \times \left(\prod_{j=1}^k\frac{\Gamma(1+2j/\beta)}{\Gamma(1+2(j-k)/\beta)}\right)
 \cos\Big(2\pi k N P_\mathrm{W}(x)+k\varphi(x,\beta)\Big)\end{multline}
for $\displaystyle
\varphi(x,\beta)=\left(1-\frac{2}{\beta}\right)\arcsin
x$.\end{corollary}

Let us consider only the very first correction to the global density:
\begin{multline}\label{AsymptHermite1st}
\sqrt{\frac{2}{N}}\rho_{N,\beta}(\sqrt{2N}x)=\rho_\mathrm{W}(x)+\mathrm{O}\left(\frac{1}{N}\right)
+\mathrm{O}\left(\frac{1}{N^{8/\beta}}\right)\\
-\frac{2}{\pi}\frac{\Gamma(1+2/\beta)}{(\pi\rho_\mathrm{W}(x))^{6/\beta-1}}\frac{1}{N^{2/\beta}}
\cos\left(2\pi N
P_\mathrm{W}(x)+\varphi(x,\beta)\right). \end{multline}
Up to a factor of order $1/N$, the dominant oscillatory terms in the Gaussian unitary and
symplectic ensembles are thus \[\sqrt{\frac{2}{N}}\rho_{N,\beta}(\sqrt{2N}x)-\rho_\mathrm{W}(x)=\begin{cases}\displaystyle\frac{-2}{\pi^3\rho_\mathrm{W}(x)^2N} \cos\left(2\pi N P_\mathrm{W}(x)
\right),&\beta=2.
\\\displaystyle
\frac{-1}{\pi \rho_\mathrm{W}(x)^{1/2}N^{1/2}}
\cos\left(2\pi N P_\mathrm{W}(x)+\frac{1}{2}\arcsin x
\right),&\beta=4,
\end{cases}\]
respectively.  A direct computation shows that the non-oscillatory
$\mathrm{O}(1/N)$ term is exactly zero when $\beta=2$.  This implies
that our result reproduces the Gaussian global densities previously
obtained in  \cite{Garoni,Kalisch}, for the unitary case,  and in
\cite{Kalisch}, for the symplectic case. \footnote{ Note however the
presence of a misprint in \cite{Kalisch} for the symplectic case.}

\begin{figure}[h]\caption{{\footnotesize Comparison of the exact density
 \eqref{ExactDensityH}, shown as a solid line,  and the asymptotic density \eqref{AsymptHermite1st}, shown as a dashed line, in the Hermite
$\beta$-ensemble for $N=7$ and  $\beta=6$.}}\label{figH1}
\begin{center}
 \begin{pspicture}(0,0)(9,7)
 \rput(5,3.5){\includegraphics[width=8cm,height=7cm]{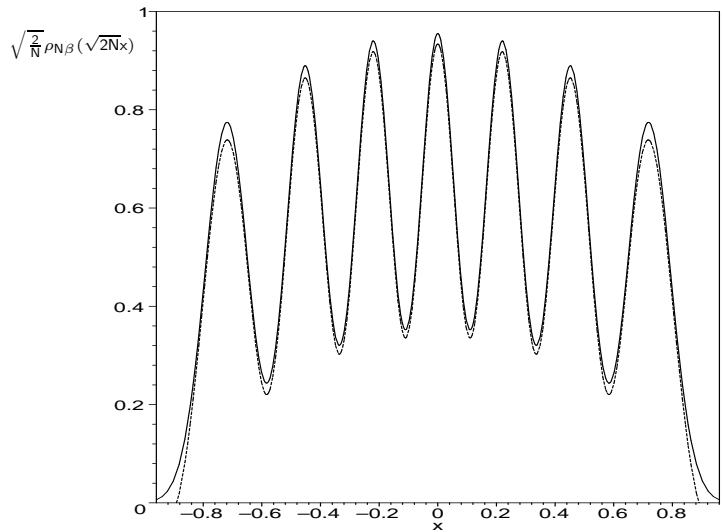}}
 \rput(0.4,6.5){{{\tiny $\mathsf{\sqrt{\frac{2}{N}}\rho_{N\beta}(\sqrt{2N}x)}$}}}
 \end{pspicture}
\end{center}
\end{figure}

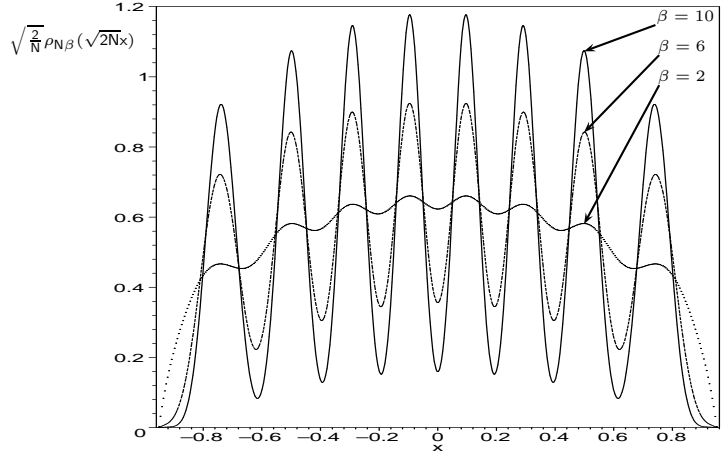
\begin{figure}[h]\caption{ {\footnotesize Asymptotic density \eqref{AsymptHermite1st} in the Hermite
$\beta$-ensemble for $N=8$ and $\beta=2,6,10$.}}\label{figH2}
\begin{center}
 \begin{pspicture}(0,0)(9,6)
 \rput(5,3){\includegraphics[width=8cm,height=6cm]{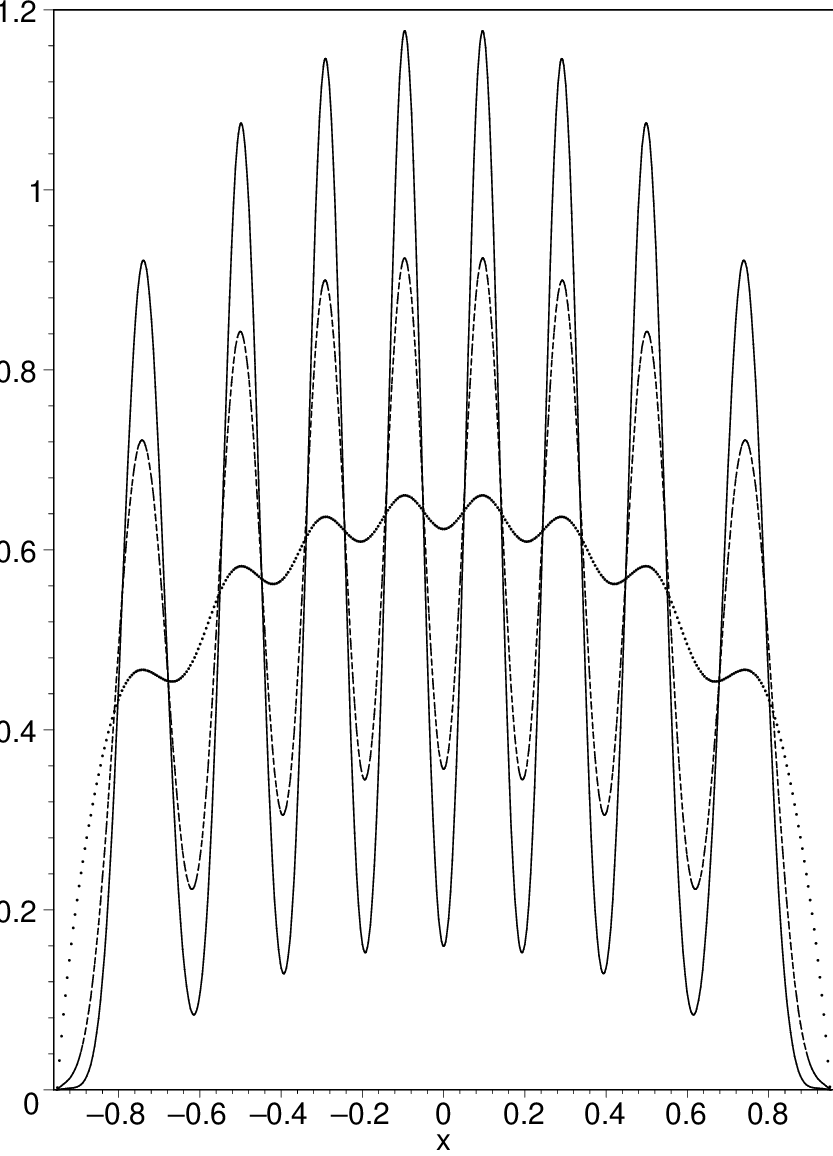}}
 \rput(0.4,5.5){{{\tiny $\mathsf{\sqrt{\frac{2}{N}}\rho_{N\beta}(\sqrt{2N}x)}$}}}
 \rput(8.55,5.8){{{\tiny ${\beta=10}$}}}\psline{->}(8.15,5.7)(7.2,5.35)
 \rput(8.5,5.4){{{\tiny${\beta=6}$}}} \psline{->}(8.2,5.3)(7.2,4.27)
  \rput(8.5,5.0){{{\tiny${\beta=2}$}}} \psline{->}(8.2,4.9)(7.2,3.05)
 \end{pspicture}
\end{center}
\end{figure}

The asymptotic expansion of the density for $\beta=6$ is numerically compared with
the exact one in Fig.~\ref{figH1}.  This picture shows that, even
for a small $N$'s,  Eq.~\eqref{ADensityH} furnishes a qualitatively
good approximation of the density in the bulk.  Fig.~\ref{figH2}
illustrates the behavior of $\rho_{N,\beta}(\sqrt{2N}x)$ when the
Dyson index varies.

\subsection{Laguerre case}  The method used to evaluate the asymptotic behavior the the Laguerre density is almost the same as the one used in the
Hermite case.  All relevant  differences  originate from the correct
contour that we must choose in   Eq.~\ref{Rdef}
\cite{ForresterLaguerre}:
 $\mathcal{C}$  starts at the point $u=1$, turns around zero in the counterclockwise  direction, and comes back to $u=1$.  In the following paragraphes,
  we briefly obtain the Laguerre version of Lemma \ref{orderedingH} and Proposition \ref{propRH}.
  We suppose that $e^{Nf(u,x)}$ is analytic everywhere, except maybe on the interval $[0,1]$.

\begin{lemma}\label{orderedingL} Let $\{\mathcal{C}_j\}$ be a set of nonintersecting  counterclockwise   contours around the origin,
all starting at $u_j=1$,  such that  $0\leq \mathrm{arg}
(u_n)\leq \ldots\leq \mathrm{arg} (u_1)\leq 2\pi$. Then
\begin{multline*}
\oint_{\mathcal{C}}du_1\cdots \oint_{\mathcal{C}}du_n\prod_{i=1}^ne^{Nf(u_i,x)} \prod_{1\leq j<k\leq
n}|u_j-u_k|^{4/\beta}= \\
n! (-1)^{n(n-1)/\beta}\,\int_{\mathcal{C}_1}du_1\cdots
\int_{\mathcal{C}_n}du_n \prod_{i=1}^ne^{N\tilde{f}(u_i,x)}
\prod_{1\leq j<k\leq n}(u_j-u_k)^{4/\beta}, \end{multline*}
where $\displaystyle \tilde{f}(u,x)=f(u,x)-\frac{2(n-1)}{\beta N}\ln u$.
\end{lemma}
\begin{proof}  We first set $|u_j|=1$, i.e., $u_j=e^{\mathrm{i}\theta_j}$.  $\mathcal{C}$  is such that $\theta_j$ goes from $0$ to $2\pi$.  The integral is completely symmetric so that we have $n!$ possible arrangements  of the type $0\leq \theta_{i_1}\leq \ldots \leq \theta_{i_n}\leq2\pi$.  We choose $0\leq \theta_{n}\leq \ldots \leq \theta_{1}\leq2\pi$.  In that case,
\[\prod_{1\leq j<k\leq
n}|u_j-u_k|^{4/\beta}=\prod_{1\leq j<k\leq
n}\left(2\sin \frac{\theta_i-\theta_j}{2}\right)^{4/\beta}.  \] The righthandside of the previous equation can be written as
\[\frac{1}{\mathrm{i}^{2n(n-1)/\beta}} \prod_{i=1}^n {u_i}^ {-2(n-1)/\beta}\prod_{1\leq j< k\leq n}(u_i-u_j)^{4/\beta} .\]
We have proved that
 \begin{multline*}
\oint_{\mathcal{C}} du_1\cdots\oint_{\mathcal{C}}  du_n \prod_{i=1}^n e^{Nf(u_i,x)} \prod_{1\leq j<k\leq
n}|u_j-u_k|^{4/\beta}=\\
 \frac{n!}{ (-1)^{n(n-1)/\beta}}
\int_{  \substack{|u_i|=1\\
0\leq \mathrm{arg}({u_n})\leq\ldots\leq\mathrm{arg}(u_{1})\leq 2\pi}} du_1\cdots du_n \prod_{i=1}^n\,  e^{N\tilde{f}(u_i,x)}
\prod_{1\leq j<k\leq n}(u_j-u_k)^{4/\beta} \end{multline*}
 The integrand is analytic everywhere but possibly on the segment $[0,1]$.  Therefore,  we  can  apply  Cauchy's theorem and deform the paths on the unit circle into any counterclockwise   contours $\mathcal{C}_i$ around zero   and starting at $u_i=1$ as long as the ordering $0\leq \mathrm{arg} (u_n) \leq\ldots\leq\mathrm{arg} (u_1)\leq 2\pi$ is satisfied.
\end{proof}

\begin{proposition} \label{propRL} Let $\tilde{f}(u,x)=f(u,x)-(2-2/\beta)N^{-1}\ln u$, where $f(u,x)$ is the function appearing
 in the definition of $R_{N,\beta}(x)$, given in Eq.~\eqref{Rdef}, and satisfying  Eqs \eqref{2sp} and \eqref{thetapm}.  Let also
$\tilde{f}_\pm=\tilde{f}(u_\pm,x)$.  Suppose moreover that the saddle points are
such that $ 0\leq \mathrm{arg}(u_+)< \mathrm{arg}(u_-)\leq 2\pi$.  Then,
\begin{multline*}R_{N,\beta}(x)=-\binom{\beta}{\beta/2}(\Gamma_{\beta/2,\beta})^2(u_+-u_-)^\beta\left(\frac{2}{NR}\right)^{\beta-1}\\
\times e^{\beta N(\tilde{f}_++\tilde{f}_-)/2}e^{\mathrm{i}(\beta-1)(\theta_++\theta_-)}
\Big[r_{N,\beta}(x)+\mathrm{O}(1/N)\Big], \end{multline*}where
\begin{multline*}
 r_{N,\beta}(x)= 1 +2\sum_{k=1}^{\lfloor\sqrt{\beta/2}\rfloor}\left[\left(\prod_{j=1}^k\frac{\Gamma(1+2j/\beta)}{\Gamma(1+2(j-k)/\beta)}\right)\right.\\
 \left.\times \frac{e^{\mathrm{i}2k^2(\theta_++\theta_-)/\beta}}{(u_+-u_-)^{4k^2/\beta}(NR)^{2k^2/\beta}}
\cos\Big(-\mathrm{i}k N
(\tilde{f}_+-\tilde{f}_-)+k(\theta_+-\theta_-)(3-2/\beta)\Big)\right]. \end{multline*}
\end{proposition}\begin{proof}We first use Lemma \ref{orderedingL}.  The remaining steps are   similar to those of Proposititon \ref{propRH}. \end{proof}

In reference \cite{ForresterLaguerre}, the  density of  the Laguerre
$\beta$-ensemble has been written in terms of  generalized
hypergeometric functions \cite{Kaneko}:
 \[ \rho_{N,\beta}( x)=N\frac{W_{a+2,\beta,N-1}}{W_{a,\beta,N}}x^{a\beta/2} e^{-\beta  x/2}{ \phantom{F}_1F_1}^{(\beta/2)}(-N+1;a+2;t_1,\ldots,t_\beta)\big|_{t_1=\ldots=t_\beta=x}.\]
 There exist integral representations of the generalized hypergeometric functions \cite{Yan}.
 For our purpose, the appropriate integral formula can be found in Chapter 11 of \cite{ForresterBook};  one easily shows that
 \begin{multline*} {\phantom{F}_1F_1}^{(\alpha)}(-B;A+1+(n-1)/\alpha;t_1,\ldots,t_n)\big|_{t_1=\ldots=t_n=x}=\\
  \frac{\mathrm{i}^{2nB}}{M_n(A,B,1/\alpha)}
 \frac{1}{2\pi\mathrm{i}}\oint_\mathcal{C}du_1\cdots \frac{1}{2\pi\mathrm{i}}\oint_\mathcal{C}du_n\prod_{j=1}^n e^{xu_j} {u_j^{-B-1}}{(1-u_j)^{A+B}}\prod_{1\leq k<l\leq n}|u_k-u_l|^{2/\alpha},\end{multline*}where $\mathcal{C}$ is as previously described and where
 \[M_n(A,B,C)=\prod_{j=1}^n\frac{\Gamma(1+A+B-C+jC)\Gamma(1+jC)}{\Gamma(1+A-C+jC)\Gamma(1+B-C+jC)\Gamma(1+C)} .\]
 This implies that the Laguerre density in the bulk can be recast in an integral of the form \eqref{Rdef}:
\begin{equation}\label{rhoRLaguerre}\rho_{N,\beta}(4Nx)=\frac{N}{(2\pi\mathrm{i})^\beta}\frac{W_{a+2,\beta,N-1}}{W_{a,\beta,N}}\frac{(4Nx)^{a\beta/2}
e^{-2\beta N
x}}{M_\beta(a+2/\beta-1,N-1,2/\beta)}R_{N,\beta}(x)\end{equation}
provided that, in Eq.~\eqref{Rdef},  $\mathcal{C}$ is a
counterclockwise closed path around the origin and starting at
$u=1$, and
\begin{equation}\label{flaguerre}f(u,x)=4xu-\ln u+\ln (1-u)
+\frac{1}{N}\left(a-2+\frac{2}{\beta}\right)\ln
(1-u).\end{equation} Neglecting   factors of order $1/N$, we see
that this function has two simple saddle points,
\[u_\pm=\frac{1}{2}\left(1\pm \mathrm{i}\sqrt{\frac{1}{x}-1}\right),   \] which satisfy $\mathrm{arg} (u_+)<\mathrm{arg}(u_-)$ only if $0<x<1$.
This implies
\[ \frac{\partial^2}{\partial
u^2}f(u,x)\big|_{u_\pm}=R e^{\mathrm{i}\phi_\pm}, \qquad R=16x^2\sqrt{\frac{1}{x}-1},\qquad \phi_\pm= \mp\frac{\pi}{2}; \] hence, the directions of steepest descent are $\theta_+=3\pi/4$ and $\theta_-=\pi/4$.  We also have
\[\tilde{f}_\pm=2x -\frac{a+4/\beta-4}{N}\ln 2\sqrt{x} \pm 2 \mathrm{i}\left(x\sqrt{\frac{1}{x}-1}-\arccos\sqrt{x}\right)\mp\mathrm{i}\frac{a}{N}\arccos\sqrt{x} .\]
The Stirling approximation readily gives
\[\frac{W_{a+2,\beta,N-1}}{W_{a,\beta,N}}=\left(\frac{\beta}{2}\right)^{1+a\beta}N^{a\beta/2}\frac{\Gamma(1+\beta/2)}{\Gamma(1+a\beta/2)\Gamma(1+(a+1)\beta/2)}\left[1+\mathrm{O}\left(\frac{1}{N}\right)\right]\]
Moreover, by using the Gauss multiplication formula, which reads
\[\prod_{j=0}^{n-1}\Gamma(z+j/n)=(2\pi)^{(n-1)/2}n^{1/2-nz}\Gamma(nz),\]one can show that
\begin{multline}\binom{\beta}{\beta/2}
\frac{(\Gamma_{\beta/2,\beta})^2}{M_\beta(a+\beta/2-1,N-1,2/\beta)}
\\ =\pi^{\beta-1}\left( \frac{\beta}{2} \right)^{-1-a\beta}\frac{\Gamma(1+a\beta/2)\Gamma(1+(1+a)\beta/2)}{\Gamma(1+\beta/2)N^{a\beta+2-\beta}}
\left[1+\mathrm{O}\left(\frac{1}{N}\right)\right].  \end{multline}
By substituting the above equations in  Proposition \ref{propRL}, we  obtain the following result.
\begin{corollary}
Let $0<x< 1$ and let $P_\mathrm{MP}(x)$ denote the (cumulative)
probability distribution associated to the  density
$\rho_\mathrm{MP}$ defined in Eq. \eqref{semicircleL}; i.e.,
\begin{equation}P_\mathrm{MP}(x)=\int_{0}^x \rho_\mathrm{MP}(t) dt=1+x\rho_\mathrm{MP}(x) -\frac{2}{\pi}\arccos x\,
.\end{equation}Then
\[ 4\rho_{N,\beta}(4N x)=\rho_\mathrm{MP}(x)r_{N,\beta}(x)+\mathrm{O}\left(\frac{1}{N}\right), \]
where
\begin{multline*}
 r_{N,\beta}(x)= 1+2\sum_{k=1}^{\lfloor\sqrt{\beta/2}\rfloor}\frac{(-1)^{k}}{(2\pi^3x^2\rho_\mathrm{MP}(x)^3 N)^{2k^2/\beta}}\\
 \times \left(\prod_{j=1}^k\frac{\Gamma(1+2j/\beta)}{\Gamma(1+2(j-k)/\beta)}\right)
 \cos\Big(2\pi k N P_\mathrm{MP}(x)+k\varphi(x,\beta)\Big), \end{multline*}
\begin{equation}\varphi(x,\beta)=\left(1-\frac{2}{\beta}\right)\frac{\pi}{2}-2a\arccos \sqrt{x}. \end{equation}\end{corollary}

Again, we look at the first correction to the global density:
\begin{multline}\label{AsymptLaguerre1st}4\rho_{N,\beta}(4N x)=\rho_\mathrm{MP}(x) +\mathrm{O}\left(\frac{1}{N}\right)+\mathrm{O}\left(\frac{1}{N^{8/\beta}}\right)\\
 -\frac{2 \Gamma(1+2/\beta)}{(2\pi^3x^2 N)^{2/\beta}{\rho_\mathrm{MP}(x)}^{6/\beta-1}}
 \cos\Big(2\pi N P_\mathrm{MP}(x)+\varphi(x,\beta)\Big)
. \end{multline}
 Hence, neglecting the possible factor $\mathrm{O}(1/N)$, the first oscillatory corrections to the global density in
 the complex and quaternionionic Wishart ensembles are
$$ {\small  4\rho_{N,\beta}(4N x)-\rho_\mathrm{MP}(x)=\begin{cases}\displaystyle \frac{-1}{\pi^3x^2{\rho_\mathrm{MP}(x)}^2 N}
 \cos\Big(2\pi N P_\mathrm{MP}(x)-2a\arccos
 \sqrt{x}\Big),&\beta=2,\\
 \displaystyle
 \frac{-1}{2^{1/2}\pi x {\rho_\mathrm{MP}(x)}^{1/2} N^{1/2}}
 \cos\Big(2\pi N P_\mathrm{MP}(x)-2a\arccos
 \sqrt{x}+\pi/4\Big),&\beta=4.
 \end{cases}}$$
 Contrary  to the Gaussian unitary ensemble, the first correction  in the complex Wishart ensemble  has a non-null (when $a\neq0$)
 correction  of order $1/N$ which is not oscillatory \cite{Garoni}.  Nevertheless, our oscillatory term is the same as the one given
 the latter reference.  The $\beta=4$ case has been  recently studied   in \cite{Garoni2}; the dominant correction is purely oscillatory
 and is equal to the term given above.

\begin{figure}\caption{{\footnotesize Comparison of the exact density \eqref{ExactDensityL}, shown as a solid line,  and the asymptotic density \eqref{AsymptLaguerre1st}, shown as a dashed line, in the Hermite
$\beta$-ensemble for $N=4$, $\beta=6$, and  $a=0,1$.}}\label{figL1}
\begin{center}
 \begin{pspicture}(0,0)(9,6)
 \rput(5,3){\includegraphics[width=8cm,height=6cm]{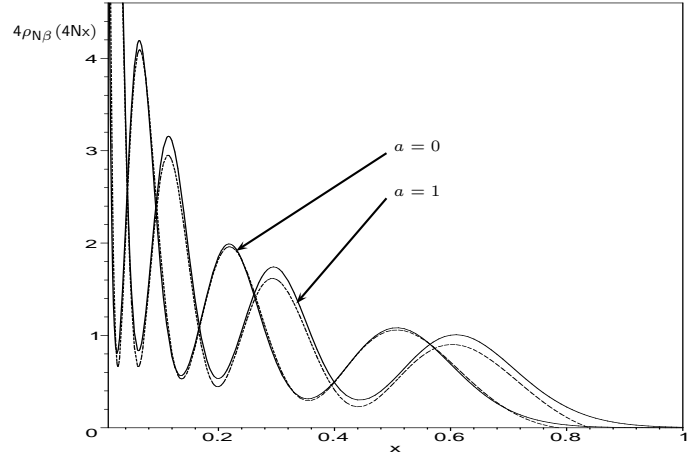}}
 \rput(0.6,5.6){{{\tiny $\mathsf{4\rho_{N\beta}(4Nx)}$}}}
  \rput(5.1,4.1){{{\tiny$\phantom{a=0}a=0$}}} \psline{->}(5,4)(3,2.7)
  \rput(5.1,3.5){{{\tiny$\phantom{a=1}a=1$}}} \psline{->}(5,3.4)(3.8,2)
 \end{pspicture}
\end{center}
\end{figure}
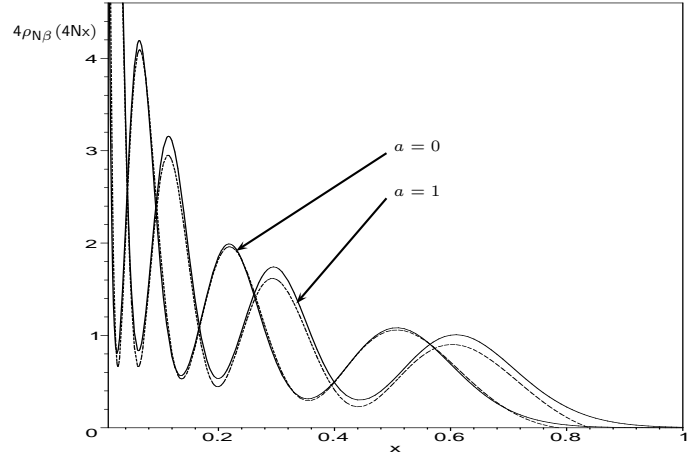

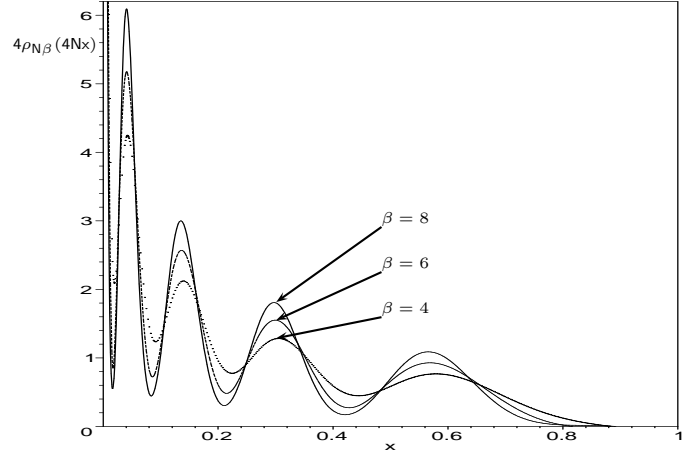
\begin{figure}\caption{{\footnotesize Asymptotic density \eqref{AsymptLaguerre1st} in the
Laguerre $\beta$-ensemble for $N=5$, $a=0$, and
$\beta=4,6,8$.}}\label{figL2}
\begin{center}
 \begin{pspicture}(0,0)(9,6)
 \rput(5,3){\includegraphics[width=8cm,height=6cm]{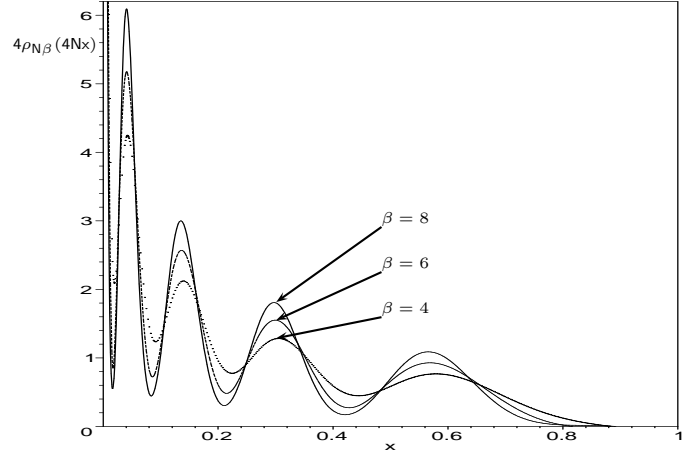}}
{\rput(0.6,5.4){ {\tiny {$\mathsf{4\rho_{N\beta}(4Nx)}$}}}}
  \rput(5,3.1){{{\tiny\phantom{$\beta=8$}$\beta=8$}}} \psline{->}(5,3)(3.6,2)
  \rput(5,2.5){{{\tiny\phantom{$\beta=8$}$\beta=6$}}} \psline{->}(5,2.4)(3.6,1.76)
  \rput(5,1.9){{{\tiny\phantom{$\beta=8$}$\beta=4$}}} \psline{->}(5,1.8)(3.6,1.52)
 \end{pspicture}
\end{center}
\end{figure}

Fig.~\ref{figL1} provides a numerical comparison between the
asymptotic and the exact expressions of the density. Clearly, the
asymptotic approximation is better for $a=0$.  Note  that the
oscillations shift to the right when $a$ increases.  In
Fig.~\ref{figL2}, we illustrate the effect of the variation of
$\beta$ on $4\rho_{N,\beta}(4N x)$ for fixed $N$ and $a$.

\section{Density at the soft-edge}
We have seen that the steepest descent method can be applied to the
scaled densities   only if $-1<x<1$ (Hermite case) or $0<x<1$
(Laguerre case). Indeed,  when the spectral parameter $x$ is outside
these intervals, the contours of integration cannot be deformed into
the steepest decent ones without transgressing the appropriate
ordering of the variables of integration.  A change of scaling is
mandatory.  The appropriate changes of variable at the soft edges
have been obtained in \cite{ForresterEdge}:  the scaled densities
$\rho_{N,\beta}(\sqrt{2N}x)$ (Hermite) and $\rho_{N,\beta}(4Nx)$ (Laguerre)
should be  replaced by $\rho_{N,\beta}(\sqrt{2N}+x/\sqrt{2N^{1/3}})$
(Hermite) and $\rho_{N,\beta}(4N+2(2N)^{1/3}x)$ (Laguerre).

Technically, these new scalings make the two simple saddle points
coalesce and become a double saddle point (or saddle point of
order two). Then, the multiple Gaussian integrals are replaced by
multiple Airy  integrals, or integrals of the Kontsevich type
\cite{Kontsevich}:
\begin{equation}\label{Kbeta}K_{n,\beta}(x):=-
\frac{1}{(2\pi\mathrm{i})^{n}}\int_{-\mathrm{i}\infty}^{\mathrm{i}\infty}dv_1\cdots
\int_{-\mathrm{i}\infty}^{\mathrm{i}\infty}dv_n\prod_{j=1}^n
e^{v_j^3/3-xv_j}\prod_{1\leq k<l\leq
n}|v_k-v_l|^{4/\beta}\,.\end{equation}We recall that the Airy
function of a real variable $x$ can be defined as follows:
\begin{equation}\label{AiryDef}\mathrm{Ai}(x)=\frac{1}{2\pi\mathrm{i}}\int_{-\mathrm{i}\infty}^{\mathrm{i}\infty}e^{v^3/3-xv}dv, \end{equation}
and as a consequence
\[\mathrm{Ai}''(x)=x\mathrm{Ai}(x).\]One readily verifies that
\[K_{1,\beta}(x)=-\mathrm{Ai}(x),\qquad
K_{2,2}(x)=2\big({\mathrm{Ai}'(x)}^2-x{\mathrm{Ai}(x)}^2
\big).\]It is worth  mentioning that the function $K_{n,2}(x)$ has
previously been studied in the context of the Gaussian unitary
ensemble \cite{Witte}.  In particular, it has been shown that
\footnote{Note however that the term $n!$ is missing in
\cite{Witte}.}
\[K_{n,2}(x)=-n!\det\left[\frac{d^{i+j-2}}{dx^{i+j-2}}\mathrm{Ai}(x)\right]_{i,j=1,\ldots,n}.\]

\subsection{Hermite case}
\begin{figure}[h]\caption{{\footnotesize
Possible contours in the integral representation of the Airy
function.}}\label{contoursAiry}
\begin{center}
 \begin{pspicture}(0,0)(7,6.5)
 \pscustom[linewidth=0pt]{ \psline(0,4.7)(3,3) \gsave
\psline(3,3)(0,1.3) \fill[fillstyle=solid,fillcolor=lightgray]
\grestore}
 \pscustom[linewidth=0pt]{ \psline(3,3)(3,6) \gsave
\psline(6,4.7)(3,3) \fill[fillstyle=solid,fillcolor=lightgray]
\grestore}
 \pscustom[linewidth=0pt]{ \psline(3,3)(6,1.3) \gsave
\psline(3,0)(3,3) \fill[fillstyle=solid,fillcolor=lightgray]
\grestore}
 \psline[linestyle=dashed]{->}(3,0)(3,6)
  \psline[linestyle=dashed]{->}(0,3)(6,3)
  \psline[linestyle=dashed]{-}(0,1.3)(6,4.7) \rput(6.4,4.9){{\small $\pi/6$}}
  \psline[linestyle=dashed]{-}(0,4.7)(6,1.3)\rput(6.4,1.2){{\small $-\pi/6$}}
  \pscurve[linewidth=1.5pt]{->}(4.8,0.8)(3.7,3)(4.8,5.2)
  \rput(5,5.4){{\small $\mathcal{A}_0$}}
  \pscurve[linewidth=1.5pt]{->}(4.4,5.4)(3,3.7)(0,3.2)
  \rput(-0.2,3.2){{\small $\mathcal{A}_1$}}
   \pscurve[linewidth=1.5pt]{->}(0,2.8)(3,2.3)(4.4,0.6)
  \rput(4.6,0.3){{\small $\mathcal{A}_{-1}$}}
     \rput(6.3,3){{\small $\Re v$}}
   \rput(3,6.3){{\small$\Im v$}}
 \end{pspicture}
\end{center}
\end{figure}
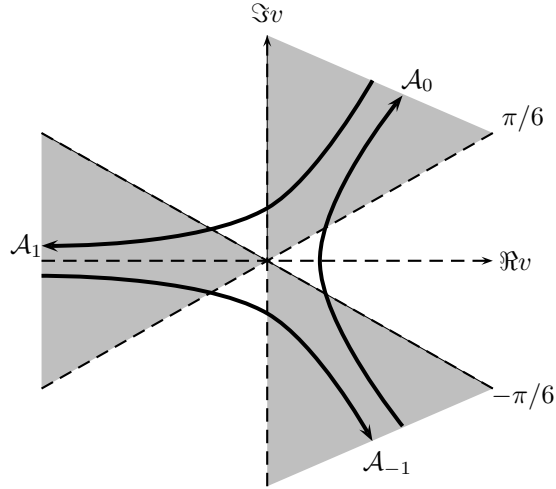

The next lemma generalizes a basic fact of the Airy function of a
complex variable~$z$; that is, the contour in Eq.~\eqref{AiryDef}
can be deformed so that
\begin{equation}\label{AiryDef2}\mathrm{Ai}(z)=\frac{1}{2\pi\mathrm{i}}\int_{\mathcal{A}_0}e^{v^3/3-zv}dv,\end{equation}
where $\mathcal{A}_0$ is a simple path going from $\infty
e^{\mathrm{i}\theta_{a}}$ to $\infty e^{\mathrm{i}\theta_{b}}$,
where $-\pi/2<\theta_a<-\pi/6$ and $\pi/6<\theta_b<\pi/2$ (see in
Fig.~\ref{contoursAiry}).

\begin{lemma}\label{orderingK}Let $\mathcal{A}_0$ be the contour described above.
Let also $\{\mathcal{V}_j\}$ denote a set of non-intersecting paths
such that $\mathcal{V}_1=\mathcal{A}_0$ and such that, for all
$j\in\{2,\ldots,n\}$, $\mathcal{V}_j$ follows $\mathcal{A}_0$ but
stop at $v_{j-1}$.  Then
\begin{equation}\label{Kbeta2}K_{n,\beta}(x):=-
\frac{n!}{(2\pi\mathrm{i})^{n}}\int_{\mathcal{V}_1}dv_1\cdots
\int_{\mathcal{V}_n}dv_n\prod_{j=1}^n e^{v_j^3/3-xv_j}\prod_{1\leq
k<l\leq n}(v_k-v_l)^{4/\beta},\end{equation}where
$-\pi<\mathrm{arg}\, v_j\leq \pi$ and where
$\mathrm{arg}\,(v_i-v_j)^{4/\beta}=0$ when both $\Im v_i=0=\Im v_j$
and $\Re v_i > \Re v_j$.\end{lemma}
\begin{proof}  We essentially proceed as in   Lemma~\ref{orderedingH}.  Firstly, we write Eq.~\eqref{Kbeta}
as  ordered integrals along the imaginary axis and remove the
absolute values.  Secondly,  the analyticity of the integrand and
the property
\begin{equation}\label{Airyregions} \lim_{R\rightarrow\infty}
\left[e^{v^3/3-xv}\right]_{v=Re^{\mathrm{i}\theta}}=0\quad
\mbox{if}\quad -\frac{\pi}{2}<\theta<
-\frac{\pi}{6},\quad\frac{\pi}{6}<\theta<\frac{\pi}{2},\quad
\mbox{or}\quad \frac{5\pi}{6}<\theta<\frac{7\pi}{6} \end{equation}
are used to deform the contour of $v_1$ into $\mathcal{A}_0$. We
finally complete the proof by exploiting the ordering of the
variables, the choice of principal branch for $(v_i-v_j)^{4/\beta}$,
and the Cauchy Theorem.
\end{proof}

\begin{proposition}\label{propRSEHermite}  The integral $R_{N,\beta}$ defined by Eqs
\eqref{Rdef} and \eqref{fhermite} satisfies
\[R_{N,\beta}\left(1+\frac{x}{2N^{2/3}}\right)=4\left(\frac{\pi}{2}\right)^\beta \frac{e^{\beta
N/2+\beta N^{1/3}x}}{2^{\beta N
}N^{\beta-2/3}}K_{\beta,\beta}(x)+\mathrm{O}\left(\frac{1}{N^{\beta-1/3}}\right)
.\]
\end{proposition}
\begin{proof}According to Lemma \ref{orderedingH}, we have that
\[R_{N,\beta}\left(1+\frac{x}{2N^{2/3}}\right)= \beta!\int_{\mathcal{C}_1}du_1\cdots
\int_{\mathcal{C}_\beta}du_\beta \prod_{i=1}^\beta
e^{Nf(u_i,1+x/2N^{2/3})} \prod_{1\leq j<k\leq
\beta}(u_j-u_k)^{4/\beta}.\]Considering $x/2N^{2/3}\ll 1$,
\[f\left(u,1+\frac{x}{2N^{2/3}}\right)=g(u,x)+\frac{x}{2(\mathrm{i}u+x)N^{2/3}}+\mathrm{O}\left(\frac{1}{N^{4/3}}\right),\]
where
\[g(u,x)=-2u^2+\ln(\mathrm{i}u+x).\]
  This functions has a double saddle point $u_0=\mathrm{i}/2$:
\[\left.\frac{\partial }{\partial u}g(u,x)\right|_{u=u_0}=0=\left.\frac{\partial^2 }{\partial u^2}g(u,x)\right|_{u=u_0},\qquad
 \left.
\frac{\partial^3 }{\partial
u^3}g(u,x)\right|_{u=u_0}=Re^{\mathrm{i}\phi_0}=16e^{-\mathrm{i}\pi/2}.\]
The directions for which the decrease of $f$ is maximum  are
determined by both conditions $\cos(3\theta_0+\phi_0)=-1$ and
$\sin(3\theta_0+\phi_0)=0$. Thus, the angles of steepest descent are
$\theta_0=-5\pi/6,-\pi/6,\pi/2$. We choose the two former and make
the following change of variables:
\[v_j=2\mathrm{i}N^{1/3} (u_j-u_0).\]This implies that
\[Nf\left(u_j,1+\frac{x}{2N^{2/3}}\right)=\frac{N}{2}-N\ln
2+N^{1/3}x+\ln
2+\frac{1}{3}v_j^3-xv_j+\mathrm{O}\left(\frac{1}{N^{1/3}}\right).\]
Let $\{\mathcal{V}_j\}$ denote the set of ordered and
non-intersecting contours of steepest descent: $\mathcal{V}_1$
starts at $\infty e^{-\mathrm{i}\pi/3}$, passes through the origin
and stops at $\infty e^{\mathrm{i}\pi/3}$; $\mathcal{V}_j$ follows
$\mathcal{V}_1$ but stops at $v_{j-1}$, where   $j=2,\ldots,\beta$.
We have proved that
\begin{multline*} R_{N,\beta}\left(1+\frac{x}{2N^{2/3}}\right)=-4\beta!\frac{e^{\beta N/2+\beta
N^{1/3}x}}{2^{\beta (N +2)}N^{\beta-2/3}\mathrm{i}^{3\beta}}\\
\times \int_{\mathcal{V}_1}dv_1\cdots
\int_{\mathcal{V}_\beta}dv_\beta \prod_{i=1}^\beta e^{v_i^3/3-xv_i}
\prod_{1\leq j<k\leq \beta}(v_j-v_k)^{4/\beta}
+\mathrm{O}\left(\frac{1}{N^{1/3}}\right).\end{multline*}
Consequently, we can apply Lemma \ref{orderingK} and the
proposition follows.
\end{proof}

 Let us go back to Eq.~\eqref{rhoRHermite} and make the
change of scaling $x\mapsto 1+x/(2N^{2/3})$:
\begin{multline*}\rho_{N,\beta}\left(\sqrt{2N}+\frac{x}{\sqrt{2N^{1/3}}}\right)\\
=\frac{1}{2}
\frac{G_{\beta,N-1}}{G_{\beta,N} \Gamma_{\beta,\beta}}(2N)^{\beta
N/2+\beta} e^{-\beta N}e^{-\beta N^{1/3}x}e^{-\beta x^2/(4N^{1/3})}
 R_{N,\beta}\left(1+\frac{x}{2N^{2/3}}\right).\end{multline*}
 Proposition
 \ref{propRSEHermite} and some manipulations directly  imply the following.
 \begin{corollary}\label{softHcoro} The density in the Hermite
 $\beta$-ensemble evaluated at soft edge is proportional to the
 an integral of  the Kontsevich type:
\begin{multline*}\frac{1}{\sqrt{2N^{1/3}}}\,\rho_{N,\beta}\left(\sqrt{2N}+\frac{x}{\sqrt{2N^{1/3}}}\right)\\
 =\frac{1}{2\pi}\left(\frac{4\pi}{\beta}\right)^{\beta/2}\frac{\Gamma(1+\beta/2)}{\prod_{j=2}^\beta
 {\Gamma(1+2/\beta)}^{-1}\Gamma(1+2j/\beta)}\,K_{\beta,\beta}(x)+\mathrm{O}\left(\frac{1}{N^{1/3}}\right).\end{multline*}
 \end{corollary}

\subsection{Laguerre case}

It is obvious from Eq.~\eqref{AiryDef2} and Fig.~\ref{contoursAiry}
that
\[\frac{1}{2\pi\mathrm{i}}\int_{\mathcal{A}_{-1}\cup\mathcal{A}_0\cup\mathcal{A}_1}e^{v^3/3-zv}dv=0\quad\Longleftrightarrow\quad
e^{-\mathrm{i}2\pi/3}\mathrm{Ai}(e^{-\mathrm{i}2\pi/3}z)+\mathrm{Ai}(z)+e^{\mathrm{i}2\pi/3}\mathrm{Ai}(e^{\mathrm{i}2\pi/3}z)=0.\]
This result can be extended to the Kontsevich type integral as
follows.

\begin{lemma}\label{contoursK}Let $\mathcal{A}_{-1},\mathcal{A}_1$ be the contours depicted in Fig.~\ref{contoursAiry}.
Let also $\{\widetilde{\mathcal{V}}_j\}$ denote a set of
non-intersecting paths such that
$\widetilde{\mathcal{V}}_1=\mathcal{A}_{-1}\cup\mathcal{A}_1$ and
such that, for all $j\in\{2,\ldots,n\}$, $\widetilde{\mathcal{V}}_j$
follows $\widetilde{\mathcal{V}}_1$ but stops at $v_{j-1}$.  Then
\begin{equation}\label{Kbeta3}K_{n,\beta}(x)=(-1)^{n+1}
\frac{n!}{(2\pi\mathrm{i})^{n}}\int_{\widetilde{\mathcal{V}}_1}dv_1\cdots
\int_{\widetilde{\mathcal{V}}_n}dv_n\prod_{j=1}^n
e^{v_j^3/3-xv_j}\prod_{1\leq k<l\leq
n}(v_k-v_l)^{4/\beta},\end{equation}where $-\pi<\mathrm{arg}\,
v_j\leq \pi$ and where $\mathrm{arg}\,(v_i-v_j)^{4/\beta}=0$ when
both $\Im v_i=0=\Im v_j$ and $\Re v_i > \Re v_j$.\end{lemma}
\begin{proof} By virtue of Eq.~\eqref{Airyregions}, it is possible to join $\mathcal{A}_{-1}$ and
$\mathcal{A}_1$. Thus, Cauchy's Theorem   implies that
$\widetilde{\mathcal{V}}_1$ can be replaced $-\mathcal{A}_0$.  The
constraint on the ordering of the variables $\{v_2,\ldots,v_n\}$ is
then considered and equivalence between Eq.~\eqref{Kbeta2} and
Eq.~\eqref{Kbeta3} follows.
\end{proof}

\begin{proposition}\label{propRSELaguerre}  The integral $R_{N,\beta}$ defined by Eqs
\eqref{Rdef} and \eqref{flaguerre} satisfies
\[R_{N,\beta}\left(1+\frac{x}{(2N)^{2/3}}\right)= (2\pi\mathrm{i})^\beta\frac{e^{2\beta N}e^{\beta
(2N)^{1/3}x}}{2^{a\beta+4/3}N^{\beta-2/3}}K_{\beta,\beta}(x)+\mathrm{O}\left(\frac{1}{N^{\beta-1/3}}\right)
.\]
\end{proposition}
\begin{proof}We  essentially follow the proof of Proposition \ref{propRSEHermite}.  By virtue of Lemma \ref{orderedingL}, we have that
\[R_{N,\beta}\left(1+\frac{x}{(2N)^{2/3}}\right)= \beta!\oint_{\mathcal{C}_1}du_1\cdots
\oint_{\mathcal{C}_\beta}du_\beta \prod_{i=1}^\beta
e^{N\tilde{f}(u_i,1+x/(2N)^{2/3})} \prod_{1\leq j<k\leq
\beta}(u_j-u_k)^{4/\beta}, \]where
\[\tilde{f}\left(u,1+\frac{x}{(2N)^{2/3}}\right)=g(u)+\frac{1}{(2N)^{2/3}}xu+\frac{1}{N}\left(a-2+\frac{2}{\beta}\right)\ln(1-u)+\frac{1}{N}\left(\frac{2}{\beta}-2\right)\ln (u),\]
for
\[g(u)=4u+\ln(1-u)-\ln(u)\,.\]The latter function possesses a double
saddle point $u_0=1/2$.  One can check that \[
 {d^3g(u)}/{du^3}|_{u=u_0} = 32e^{\mathrm{i}\pi},\] so  the steepest descent angles are $\theta_0=0,2\pi/3,4\pi/3$.
 The contour of $u_1$ is chosen such that:
 (1) it approaches $u_0$ by following the real axis in the negative direction;
 (2) it leaves the saddle point with an angle   $2\pi/3$; (3) it turns around
 the origin in the positive direction; (4) it comes back to the $u_0$
 with an argument of $4\pi/3$; (5) it finally leaves this point and
 reaches the point $u=1$ by following the real axis.
 The ordering of the variables around the origin implies moreover that $u_i$ follows $u_1$ but stops at
 $u_{i-1}$. When $N$ is large, step (3) is irrelevant and  the steepest
 descent contour brakes into two disjoints paths, namely (1)-(2) and (4)-(5).
 Now we
 set
\[\tilde{v}_j=2(2N)^{1/3} (u_0-u_j), \]which means that
\[
N\tilde{f}\left(u_j,1+\frac{x}{(2N)^{2/3}}\right)=2N+(2N)^{1/3}x+\left(4-\frac{4}{\beta}-a\right)\ln
2+\frac{1}{3}{ \tilde{v}_j }^3
-x\tilde{v}_j+\mathrm{O}\left(\frac{1}{N^{1/3}}\right).\] When
$N\rightarrow\infty$, the  contours of the variables
$\{\tilde{v}_j\}$ that give the major contribution to the integral,
denoted by $\{\widetilde{\mathcal{V}}_j\}$, behave as follows:
$\widetilde{\mathcal{V}}_1$ is the union of the path that begins at
$-\infty$, passes close to the origin and ends at $\infty
e^{-\mathrm{i}\pi/3}$ together with the path that starts at $\infty
e^{\mathrm{i}\pi/3}$, goes near the origin and stops at $-\infty
e^{-\mathrm{i}\pi/3}$ ; $\widetilde{\mathcal{V}}_j$ follows
$\widetilde{\mathcal{V}}_1$ but stops at $v_{j-1}$, where
$j=2,\ldots,\beta$. Therefore,
\begin{multline*}
R_{N,\beta}\left(1+\frac{x}{(2N)^{2/3}}\right)=\beta!\frac{ e^{2\beta N}e^{\beta
(2N)^{1/3}x}}{2^{a\beta+4/3}N^{\beta-2/3}}\\
\times \int_{\widetilde{\mathcal{V}}_1}d\tilde{v}_1\cdots
\int_{\widetilde{\mathcal{V}}_\beta}d\widetilde{v}_\beta
\prod_{i=1}^\beta e^{{\tilde{v}_i}^3/3-x\tilde{v}_i} \prod_{1\leq
j<k\leq \beta}(\tilde{v}_j-\tilde{v}_k)^{4/\beta}
+\mathrm{O}\left(\frac{1}{N^{1/3}}\right).\end{multline*}
Lemma~\ref{contoursK} finally  provides the sought for result.
\end{proof}

We apply the last proposition to the scaled expression of the Laguerre density given in \eqref{rhoRLaguerre}:
\begin{multline*}\rho_{N,\beta}\left(4N+2(2N)^{1/3}x \right)\\
=\frac{N}{(2\pi\mathrm{i})^\beta}\frac{W_{a+2,\beta,N-1}}{W_{a,\beta,N}}\frac{(4N)^{a\beta/2}
e^{-2\beta N}e^{-\beta
(2N)^{1/3}}}{M_\beta(a+2/\beta-1,N-1,2/\beta)}
 R_{N,\beta}\left(1+\frac{x}{(2N)^{2/3}}\right).\end{multline*}
 Minor manipulations and use of Stirling's approximation give the sought limiting soft edge density, which is identical to that obtained in Corollary \ref{softHcoro} for the Hermite $\beta$-ensemble.
 \begin{corollary}The density in the Laguerre
 $\beta$-ensemble evaluated at the soft edge is proportional to the
 an integral of  the Kontsevich type:
\begin{multline*} 2(2N)^{1/3}\rho_{N,\beta}\left(4N+2(2N)^{1/3}x\right)\\
= \frac{1}{2\pi}\left(\frac{4\pi}{\beta}\right)^{\beta/2}\frac{\Gamma(1+\beta/2)}{\prod_{j=2}^\beta
 {\Gamma(1+2/\beta)}^{-1}\Gamma(1+2j/\beta)}\,K_{\beta,\beta}(x)+\mathrm{O}\left(\frac{1}{N^{1/3}}\right).\end{multline*}\end{corollary}

\section{Asymptotics of the Kontsevich type integral}  Here we
consider the leading order of $K_{\beta,\beta}(x)$ when
$x\rightarrow\pm \infty$.  This allows to match the soft edge
density with the bulk density expanded about the edge.

\begin{proposition}\label{Kbetaxlarge}When $x$ is large and positive
\[K_{\beta,\beta}(x)=\frac{\Gamma_{\beta,\beta}}{(2\pi)^\beta}
\frac{e^{-\frac{2\beta}{3}
x^{3/2}}}{x^{3\beta/4-1/2}}+\mathrm{O}\left(\frac{
1}{x^{3\beta/4+1}}\right).\]\end{proposition}
\begin{proof}Following the discussion in the proof of Proposition \ref{propRSEHermite},  we first change the contours in the Kontsevich like integral:
\[ K_{\beta,\beta}(x)=-\frac{\beta!}{(2\pi\mathrm{i})^\beta}\int_{\mathcal{V}_1}dv_1\cdots \int_{\mathcal{V}_\beta}dv_\beta
\prod_{i=1}^\beta e^{v_j^3/3-xv_j} \prod_{1\leq j<k\leq
\beta}(v_j-v_k)^{4/\beta},\]where $\{\mathcal{V}_j\}$ is such that
$\mathcal{V}_1$ goes from  $\infty e^{-\mathrm{i}\theta}$ to $\infty
e^{\mathrm{i}\theta}$, passing  through the point $\sqrt{x}$, and
such that $\mathcal{V}_j$ goes from  $\infty e^{-\mathrm{i}\theta}$
to $v_{j-1}$ for all $j=2,\ldots,\beta$ and $\pi/6<\theta<\pi/2$. We
now set
\[w_j=x^{1/4}(v_j-x^{1/2})e^{-\mathrm{i}\pi/2}\,.\]Thus,
\[K_{\beta,\beta}(x)= \frac{\beta!}{(2\pi)^\beta}\frac{e^{-2\beta x^{3/2}/3}}{x^{3\beta/4-1/2}}
\int_{\mathcal{W}_1}dw_1\cdots \int_{\mathcal{W}_\beta}dw_\beta
\prod_{i=1}^\beta e^{-w_j^2-\mathrm{i}w_j^3/3x^{3/4}} \prod_{1\leq
j<k\leq \beta}(w_j-w_k)^{4/\beta},\]where
$\mathcal{{W}}_j=e^{-\mathrm{i}\pi/2}\mathcal{{V}}_j$. By virtue of
Lemma \ref{orderedingH}, we can write
\[K_{\beta,\beta}(z)= \frac{1}{(2\pi)^\beta}\frac{e^{-2\beta
x^{3/2}/3}}{x^{3\beta/4-1/2}} \int_{-\infty}^\infty dw_1\cdots
\int_{-\infty}^\infty dw_\beta \prod_{i=1}^\beta
e^{-w_j^2+\mathrm{O}(x^{-3/4})} \prod_{1\leq j<k\leq
\beta}|w_j-w_k|^{4/\beta}.\]Note that the term
$\mathrm{O}(x^{-3/4})$ is odd in $w_j$.   As explained in the proof
of Proposition \ref{rhoRHermite}, this implies that the actual
correction to the integral is of order $x^{-3/2}$.  We finally
obtain the desired expression by comparing the last equation with
Eq.~\eqref{defGamma}.
\end{proof}

Applying Proposition \ref{Kbetaxlarge}, the next result gives the behavior of the density when the spectral
parameter leaves the bulk.
\begin{corollary}\label{DensitySoft}
Let $\sigma(x)$ denote the density evaluated at the soft edge:
\[\sigma(x)=\begin{cases}
\displaystyle \lim_{N\rightarrow\infty}\frac{1}{\sqrt{2N^{1/3}}}\rho_{N,\beta}\left(\sqrt{2N}+\frac{x}{\sqrt{2N^{1/3}}}\right),&\mbox{(Hermite)}\\
\displaystyle
\lim_{N\rightarrow\infty}2(2N)^{1/3}\rho_{N,\beta}\left(4N+2(2N)^{1/3}x\right),&\mbox{(Laguerre)}.\end{cases}\]
Then, as $x\rightarrow\infty$,
 \[ \sigma(x)=\frac{1 }{2\pi}\frac{\Gamma(1+\beta/2)}{(4
\beta)^{\beta/2}}\frac{e^{-\frac{2\beta}{3}
x^{3/2}}}{x^{3\beta/4-1/2}}+\mathrm{O}\left(\frac{
1}{x^{3\beta/4+1}}\right).\]
\end{corollary}

When the density is evaluated at points inside the bulk but close to
the edge, we should observe both decrease and oscillation (see
Fig.~\ref{figH1}-\ref{figL2}). This is confirmed in next
paragraphs.

\begin{proposition}\label{Ksouth} Let  $x=-|x|$.  When $|x|$ is large,
\[K_{\beta,\beta}(x)=\frac{(\Gamma_{\beta/2,\beta})^2}{\pi^\beta}\binom{\beta}{\beta/2}
  \sqrt{|x|}\,  k_{x,\beta}  +\mathrm{O}\left(\frac{1}{x^{5/2}}\right) , \]
  where
\[k_{x,\beta}=1+2\sum_{k=1}^{\lfloor\sqrt{\beta/2}\rfloor}\frac{(-1)^{k}}{2^{6k^2/\beta}|x|^{3k^2/\beta}}\left(\prod_{j=1}^k\frac{\Gamma(1+2j/\beta)}{\Gamma(1+2(j-k)/\beta)}\right)
 \cos\left(\frac{4k}{3}|x|^{3/2}-\frac{\pi}{2}k\left(1-\frac{2}{\beta}\right)\right).\] \end{proposition}
\begin{proof}
By   rescaling  the variables in Eq. \eqref{Kbeta}, we get
\[K_{\beta,\beta}(x)=-\frac{|x|^{3\beta/2-1}}{(2\pi\mathrm{i})^\beta}\int_{-\mathrm{i}\infty}^{\mathrm{i}\infty}dv_1\cdots
\int_{-\mathrm{i}\infty}^{\mathrm{i}\infty}dv_\beta\prod_{j=1}^\beta
e^{|x|^{3/2}f(v_j)}\prod_{1\leq k<l\leq \beta}|v_k-v_l|^{4/\beta},\]
where
\[f(v)=\frac{1}{3}v^3+ v.\]
The function $f$ possesses two simple saddle points, namely,
$v_\pm=e^{\pm\mathrm{i}\pi/2}$.  We have $f_\pm=f(v_\pm)=\pm
2\mathrm{i}/3$ and $f''(v_\pm)=2e^{\pm\mathrm{i}\pi/2}$; whence the
angle of steepest descent are $\theta_\pm=\pi/2\mp \pi/4$.  The
remainder  of the proof is a straightforward application of
  Proposition~\ref{propRH}.\end{proof}

\begin{corollary}\label{DensitySoftOsc} Let $\sigma(x)$ be the quantity defined in Lemma~\ref{DensitySoft}.  When $x\rightarrow-\infty$,  we have
\begin{multline*}
\pi \sigma(x)=\sqrt{|x|}
-\frac{\Gamma(1+\beta/2)}{2^{6/\beta-1}|x|^{3/\beta-1/2}}\cos\left(\frac{4}{3}|x|^{3/2}-\frac{\pi}{2}\left(1-\frac{2}{\beta}\right)\right)
\\+\mathrm{O}\left(\frac{ 1}{|x|^{5/2}}\right)+\mathrm{O}\left(\frac{
1}{|x|^{6/\beta-1/2}}\right).\end{multline*}
\end{corollary}

The previous result can be obtained directly from the asymptotic
density in the bulk of the Hermite (or Laguerre) $\beta$-ensemble.
Indeed, the change of variable $x\mapsto1-|x|/(2N^{2/3})$ in
Eq.~\eqref{ADensityH}  and the development of this expression for
$N^{-1/3}|x|\ll 1$ reclaims Corollary~\ref{DensitySoftOsc}. However,
it is impossible to derive Corollary~\ref{DensitySoft} from
Eq.~\eqref{ADensityH} by such an expansion.  Note finally that
Corollaries~\ref{DensitySoft}-\ref{DensitySoftOsc} imply that the
density at the soft edge of the Laguerre $\beta$-ensemble is
independent of $a$ when both $N$ and $|x|$ are large.

\section{Concluding
remarks}

The aim of the article was to determine the large-$N$ asymptotic
expansion of the density in the Hermite and Laguerre
$\beta$-ensembles when $\beta\in2\mathbb{N}$.

We have shown that  the first correction to the global density is
purely oscillatory when $\beta>2$ and is of order $N^{2/\beta}$. In
the  Hermite ensemble of $N\times N$ random matrices, the density
contains  $N$ peaks; the greater is $\beta$ and the higher are the
oscillations. The influence of the Dyson parameter on the
oscillations is the same in the Laguerre ensemble. However, the
density in the latter ensemble contains $N-1$ summits and a (delta)
divergence at the origin.

These results agree with the large-$\beta$ asymptotic analysis
realized recently in \cite{DumitriuAsympto}.  More precisely, it has
been proved that for $\beta\rightarrow\infty$, the density in the
bulk of the Hermite ensemble can be written as a sum of $N$ Gaussian
distributions centered at the zeros of an Hermite polynomial of
degree $N$ (and similarly for the Laguerre case).  These conclusions
are, of course, coherent with the log-gas analogy presented in Section 1.
 Note that no constraints on $\beta$ are imposed in
\cite{DumitriuAsympto}. Consequently, we may surmise that our
asymptotic formulas \eqref{AsymptHermite1st} and \eqref{AsymptLaguerre1st}
 are valid for any real $\beta$ greater that $2$, though the general method to prove this is still missing.

 We have also shown that the density of the Hermite and Laguerre
 ensembles are both proportional to a Kontsevich like integral $K_{\beta,\beta}(x)$ when evaluated about the edges of the
 spectrum.  Although the exact densities of the Hermite and Laguerre ensembles are quite different, the asymptotic analysis of
 $K_{\beta,\beta}(x)$ has revealed that they approach the same function in the soft edge scaling, thus verifying the expected universality.  The Kontsevich like integral itself is a special function generalizing the Airy integral and, as such, is worthy for independent study.

\vspace{0.6cm}

\noindent {\bf Acknowledgments.}  The work of P.J.F. has been
supported by the Australian Research Council.
 P.D. is grateful to the Natural Sciences and Engineering Research Council of Canada for a postdoctoral  fellowship.

\end{document}